\documentclass[12pt,showpacs,twocolumn, preprintnumbers,aps,pre, reprint,superscriptaddress]{revtex4-1}
\usepackage{graphicx}
\usepackage{siunitx}
\usepackage{subfigure}
\usepackage{color}
\usepackage{amsmath}
\usepackage{bm}
\usepackage[linktocpage,colorlinks=true,linkcolor=blue,citecolor=blue,breaklinks=true]{hyperref}
\usepackage{breakcites}

\begin{document}

%\preprint{}

\title{Minimal Data Fidelity for Detection of Stellar Features or Companions}

\author{Sahil Agarwal}
\affiliation{Program in Applied Mathematics, Yale University, New Haven, USA}
\email[]{sahil.agarwal@aya.yale.edu}

\author{J. S. Wettlaufer}
\affiliation{Program in Applied Mathematics, Yale University, New Haven, USA}
\affiliation{Nordita, Royal Institute of Technology and Stockholm University, SE-10691 Stockholm, Sweden}
\email[]{john.wettlaufer@yale.edu}

\date{\today}

\begin{abstract}
Technological advances in instrumentation have led to an exponential increase in exoplanet detection and scrutiny of stellar features such as spots and faculae. 
While the spots and faculae enable us to understand the stellar dynamics, exoplanets provide us with a glimpse into stellar evolution. While the ubiquity of noise (e.g., telluric, instrumental, or photonic) is unavoidable, combining this with increased spectrographic resolution compounds technological challenges.  
To account for these noise sources and resolution issues, we use a temporal multifractal framework to study data from the SOAP 2.0 tool, which simulates a stellar spectrum in the presence of a spot, a facula or a planet. Given these controlled simulations, we vary the resolution as well as the signal-to-noise (S/N) ratio to obtain a lower limit on the resolution and S/N required to robustly detect features. We show that a spot and a facula with a 1\% coverage of the stellar disk can be robustly detected for a S/N (per pixel) of 35 and 60 respectively, for any {spectral resolution above 20,000}, while 
a planet with a radial velocity (RV) of 10 ms$^{-1}$ can be detected for a S/N (per pixel) of 600. Rather than viewing noise as an impediment, our approach uses noise as a source of information. 
\end{abstract}

\pacs{}

\maketitle

\section{Introduction} 

The study of spots and faculae on Sun-like stars underlies a key feature of exoplanet research, which has seen substantial growth in the past decade, motivated by fundamental biological and physical questions \citep{Lanza:2008aa, Czesla:2009aa, Zellem:2017aa}. Theoretical studies substantially precede observational evidence \citep{Schatten:1978aa, Zwaan:1978aa, Hoyt:1aa, Frohlich:2004aa, Foukal:2006aa, Shapiro:2016aa} which, due to the advent of large telescopes and satellites, has systematically emerged during the past several decades. Understanding the properties of a star along with its variability (and companions if present), is foundational to such studies.
%Whether the main aim is to study the formation of planetary systems, galaxies or the universe itself, properties of a star along with its variability and companions if present, provide an insight into such studies. 
%The recent decade has also sought to find evidence of extra-terrestrial life, which led to rise in exoplanet detections.  

The most easily observed surface features of stellar variability are spots (dark regions) or faculae (light regions).   The general origin of these features is associated with magnetohydrodynamic and thermal convective processes, although their precise origin is the subject of ongoing study \cite[See e.g.,][and references therein]{Berdyugina:2005aa, Shravan, Choudhuri, Jorg}.
%Star spots are caused due to magnetic fluctuations on the stellar surface, whereas faculae are the bright spots on the surface of the star. 
%There have been many theories and attempts to model these stellar structures \cite[See e.g.,][and references therein]{Berdyugina:2005aa} in order to improve our understanding of stellar dynamics. 
Although these stellar features have enabled us to study many processes, such as differential rotation and stellar evolution, with respect to the detection of exoplanets, 
they have also been viewed as obstructive \emph{noise}.  This latter perspective derives from the fact that spots and faculae can mimic the presence of an exoplanet, due to both changes in the radial velocity of the star and by masquerading as a transit event.
%These stellar features have enabled us to study processes such as differential rotation, stellar evolution, and many other properties. 
%Although these features have importance for the reasons just described, they have also been associated with \emph{noise}. 
%This \emph{noise} is what has been tried to be \emph{removed} when detecting exoplanets, as these spots and faculae can mimic the presence of an exoplanet due to both changes in radial velocity of the star as well as simulating a transit event.

Detection of exoplanets is major accomplishment in physical science \citep{Wolszczan:1992aa, Mayor:1995aa, Borucki:2010aa}. The two most common methods of detecting exoplanets are the Radial Velocity Method and the Transit Method \cite[e.g.,][]{Aigrain:2004aa, Fischer:2014, Fischer:2016aa}. The radial velocity method measures the doppler shift in the stellar spectrum, which is related to the relative center of mass motion of the star-planet system. This is a practical approach only when a Jupiter sized planet has an orbit at the distance of Mercury, and hence such exoplanets are typically called Hot Jupiters.  Although in principle the method is simple, false detections commonly occur due to overfitting \citep{Dumusque:2012aa, Rajpaul:2016aa}.  Moreover, due to the high precision of the measurements required to detect the movement of the star, the method is very sensitive to noise \citep{Struve:1952aa, Lovis:2010aa, Lanza:2011kt, Aigrain:2012rz}. Therefore, the approach requires the removal of noise from the ``raw'' signal, the sources of noise including stellar surface activity (granulation, stellar spots, faculae), long term stellar activity, instrumental noise, and atmospheric/telluric noise. The transit method is based upon the decrease in the intensity flux coming from a host star as a planet passes between the star and the observer. Because this decrease in flux is typically about 0.5-2\%, this method is also very sensitive to noise.  Additionally, intrinsic physical phenomenon, in particular spots and faculae, can easily mimic a planet, leading to a high false detection rate \citep{Desort:2007aa, Oshagh:2014aa, Ligi:2015aa, Korhonen:2015aa}. 

\begin{figure*}[htbp!]
        (a)\includegraphics[trim = 0 0 0 0, clip, width = 0.45\textwidth]{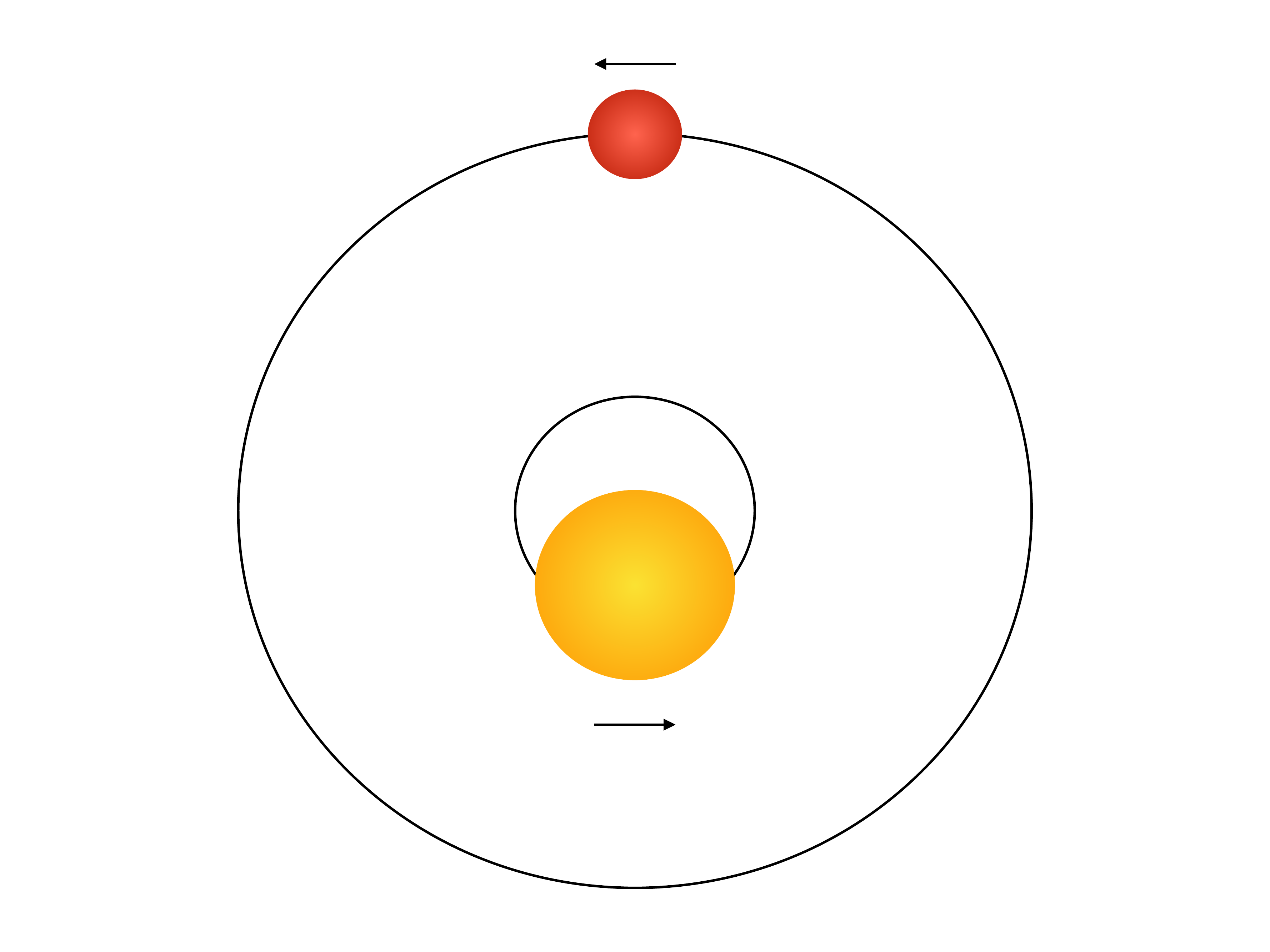}	  	
        (b)\includegraphics[trim = 0 0 0 0, clip, width = 0.45\textwidth]{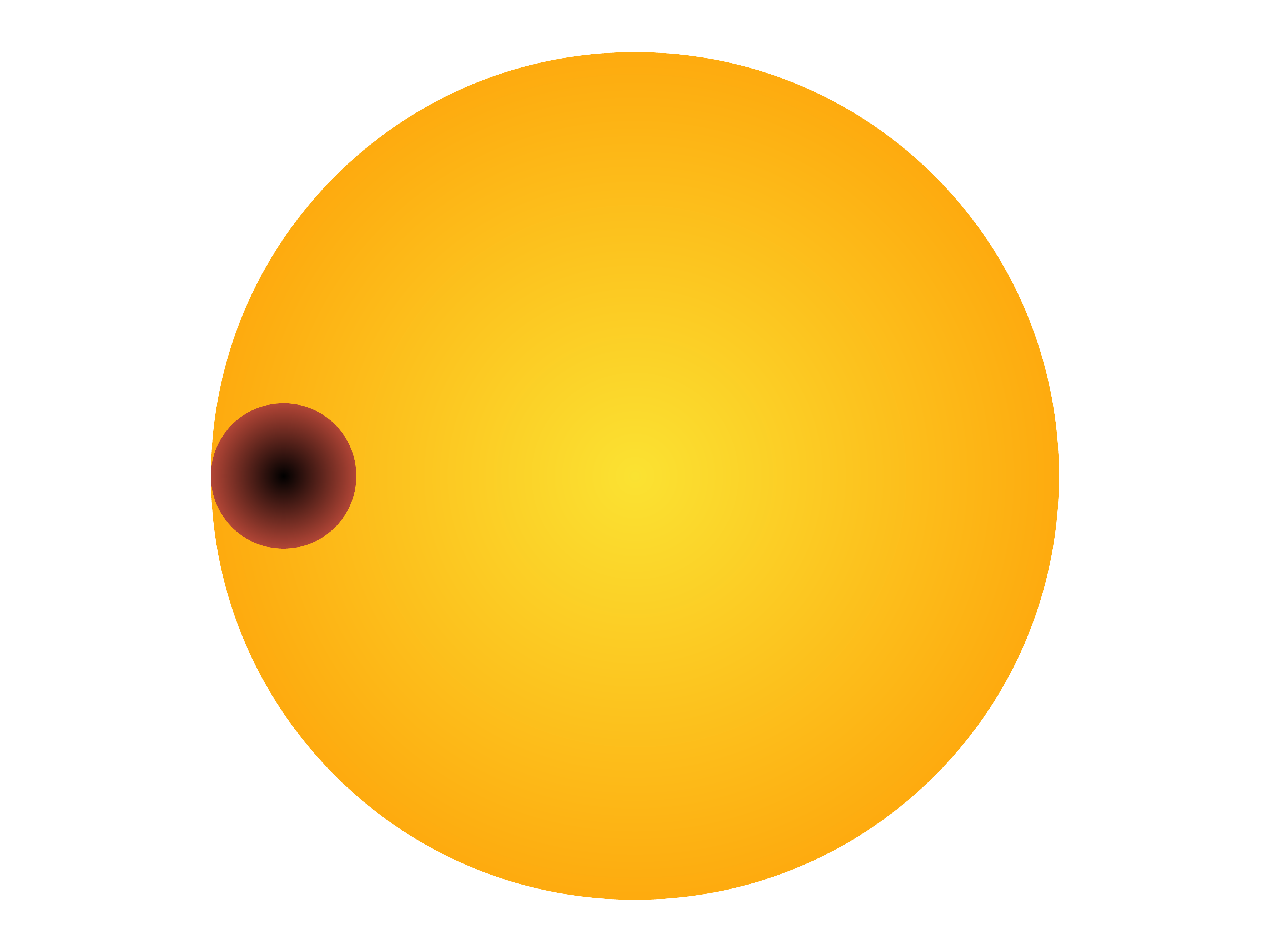}
    \caption{(a) Plan view of a planetary system orbiting the system's center of mass, i.e., both the planet and the star gravitationally interact, resulting in a Doppler effect from the light being observed from the star. (b) A spot on the stellar surface, as the star rotates about its axis.}
    \label{fig:Cartoon}
\end{figure*}

The rotation of a ``clean'' star, i.e., a star without surface spots or faculae, will have a net zero shift in the spectrum. Namely, the blue shift caused by the incoming half of the stellar surface is balanced by the red shift of the outgoing half. However, this balance can be broken due to the presence of spots or faculae, thereby producing a net radial velocity curve for the star.  Because this radial velocity can mimic the presence of a planet, it is important to be able to distinguish between the two signals. 

%Whether it is looking for stellar features or exoplanets, 
Regardless of whether one is studying stellar features or detecting exoplanets, one seeks data with high resolution and large Signal-to-Noise ratio (S/N). Data degradation can arise either due to unavoidable instrumental problems or noise sources, such as telluric contamination or light from a nearby star.  This has often led to removal of datasets from analyses so that only \emph{clean} datasets are examined. Here we quantitatively examine how noisy the data can be and yet still contain sufficient statistical information about the stellar features present. However, rather than removing the noise from the signal, we treat it as a source of information. Central to our approach is to utilize the separation of timescales of the noise arising from different physical phenomena.

%This has often led to removal of such datasets from the analysis and only \emph{clean} datasets are examined. Here we study how noisy can the data be and still contain sufficient statistical information about the stellar feature present in the \emph{clean} data. To this effect, rather than removing the noise from the signal, we treat it as a source of information. Because the timescales over which different noise processes act are different, our main approach will be to utilize this separation of timescales of different physical phenomena.

\begin{figure*}[htbp!]     
 	 (a)\includegraphics[trim = 0 150 0 0, clip, width = 0.45\textwidth]{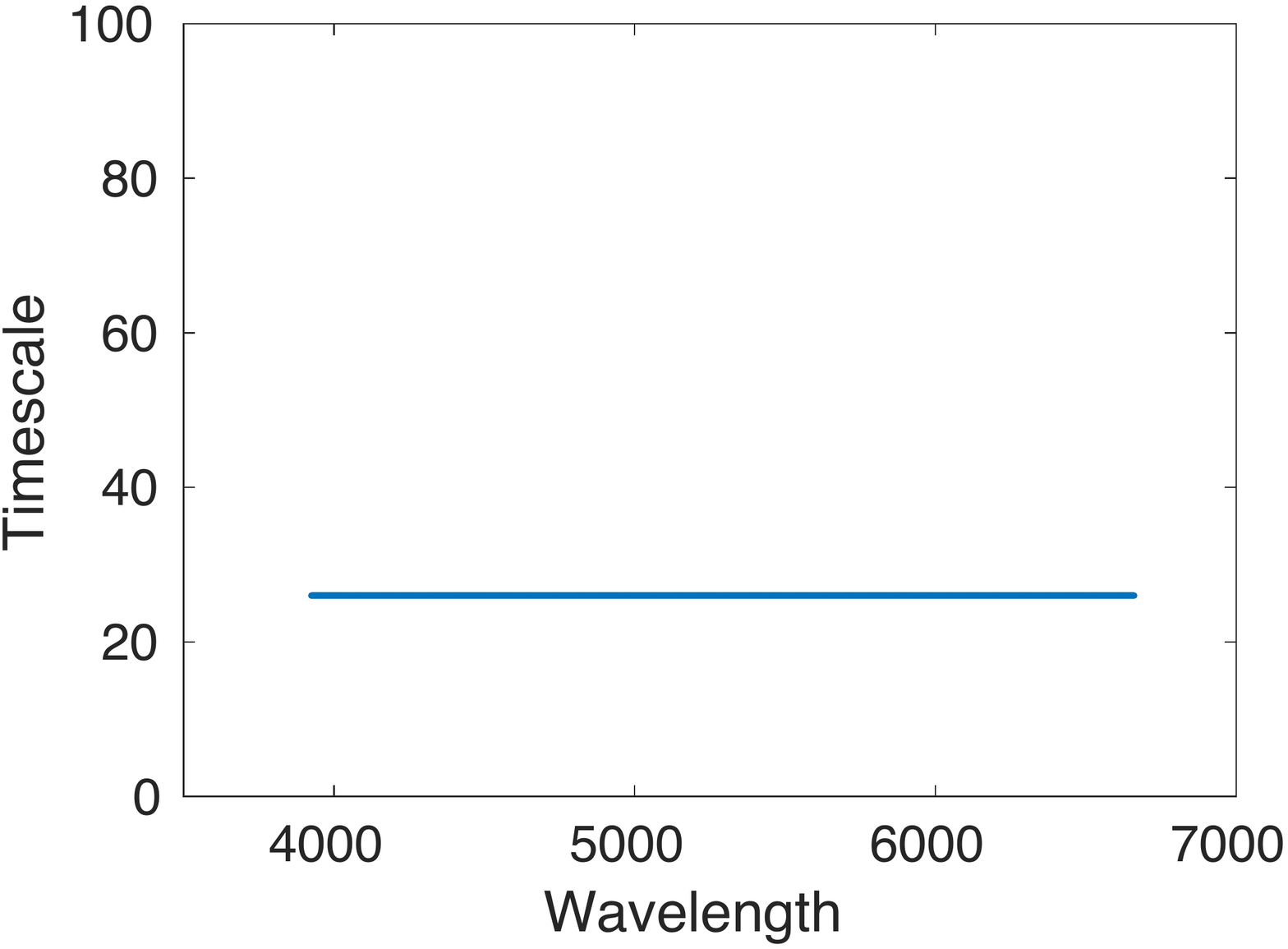}   	
      %  (a)\includegraphics[trim = 0 0 0 0, clip, width = 0.45\textwidth]{Figs/spot_1pc_orig.pdf}	  	
        %(d)\includegraphics[trim = 0 0 0 0, clip, width = 0.45\textwidth]{Figs/spot_5pc_orig.pdf}
        (b)\includegraphics[trim = 0 150 0 0, clip, width = 0.45\textwidth]{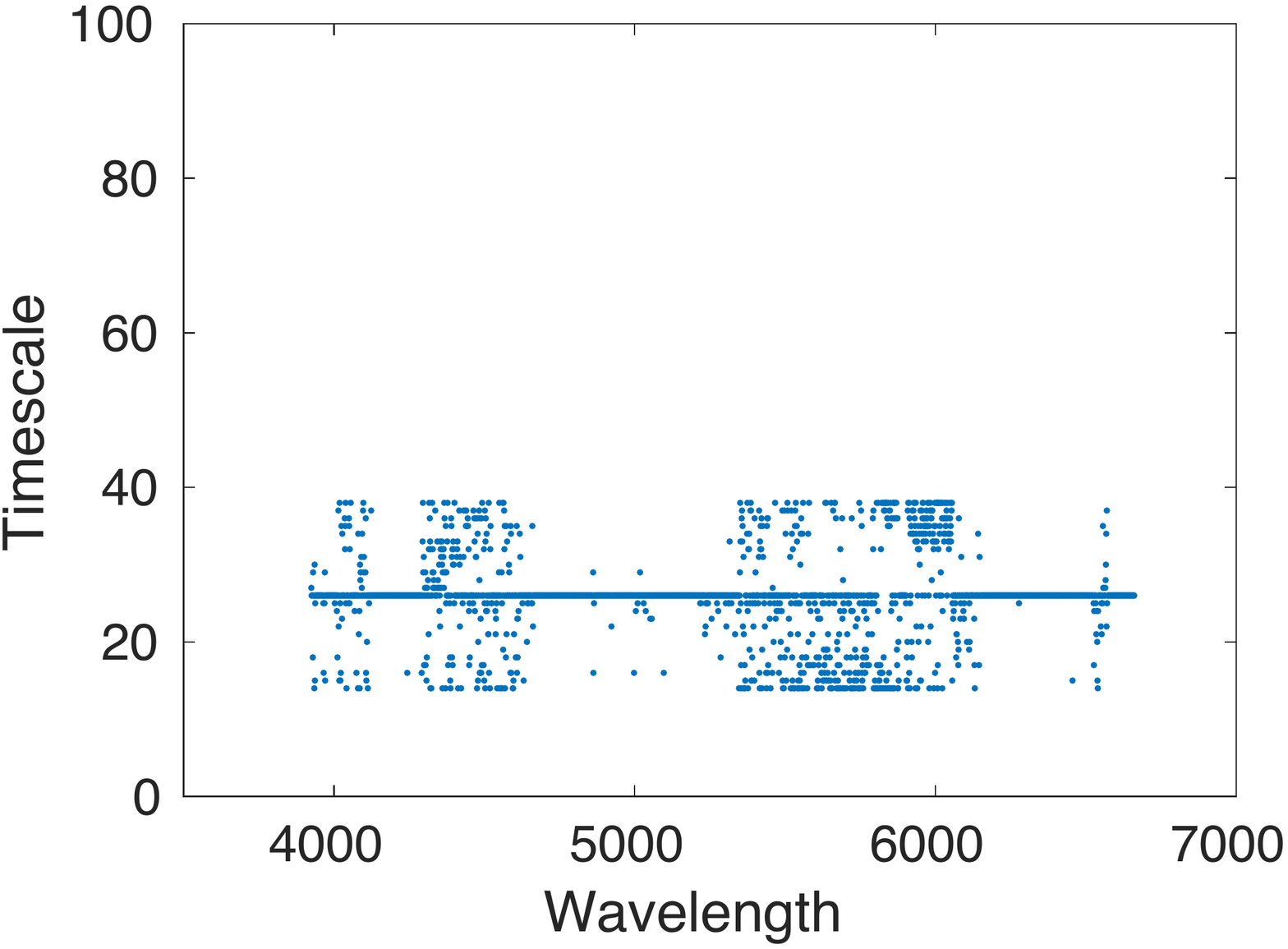}	  	
        %(b)\includegraphics[trim = 0 0 0 0, clip, width = 0.45\textwidth]{Figs/facula_5pc_orig.pdf}
        (c)\includegraphics[trim = 0 150 0 120, clip, width = 0.45\textwidth]{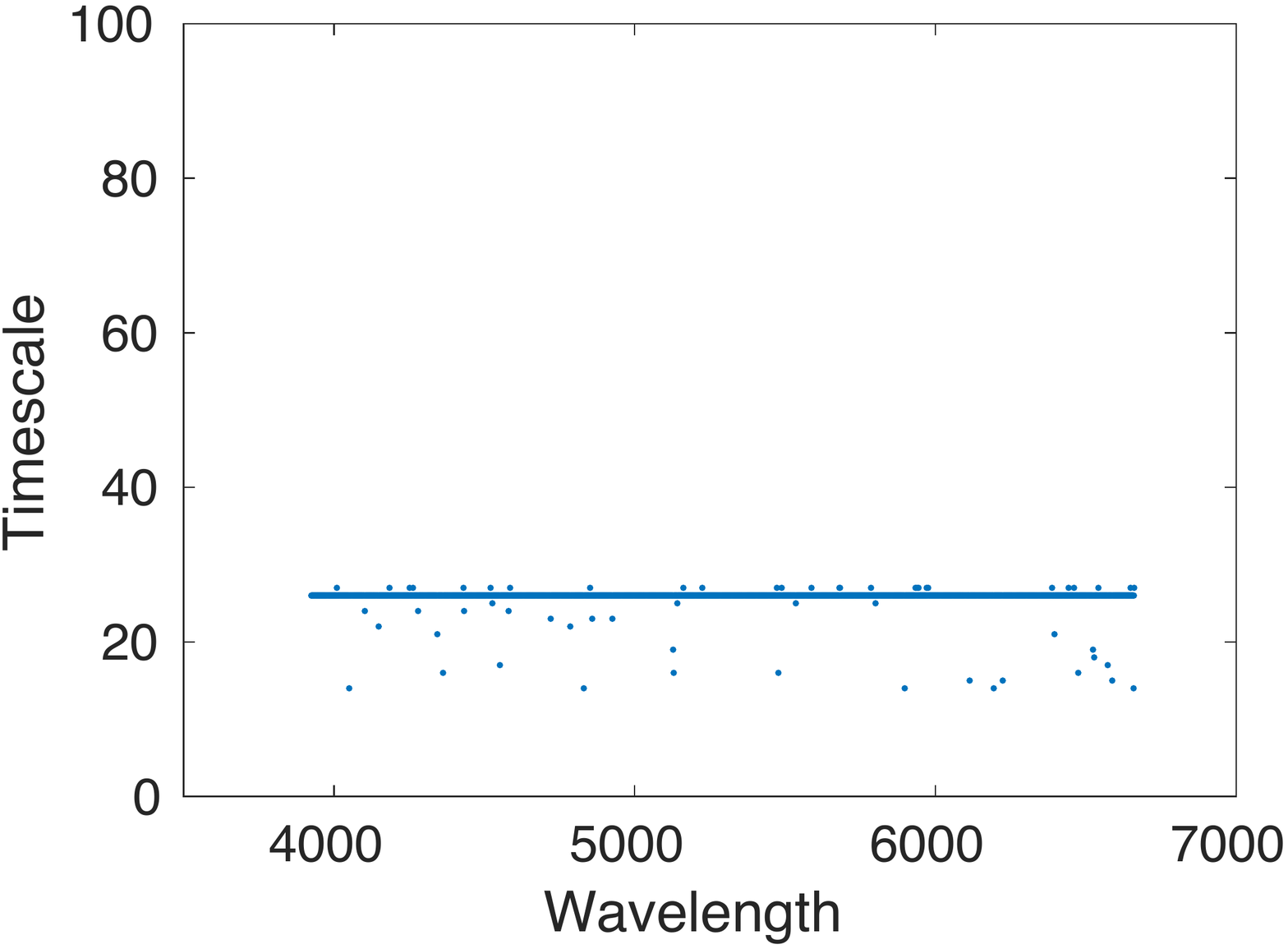}
        %(f)\includegraphics[trim = 0 0 0 0, clip, width = 0.45\textwidth]{Figs/planet_50ms_orig.pdf}
    \caption{Crossover timescales  for the wavelengths (in \si{\angstrom}) obtained by MF-TW-DFA with the original data for (a) the spot (with coverage 1\%), (b) facula (with coverage 1\%) and (c) planet (with RV 10 ms$^{-1}$).}
    \label{fig:CO_Orig}
\end{figure*}

\begin{figure*}[htbp!]
        (a)\includegraphics[trim = 0 150 0 120, clip, width = 0.45\textwidth]{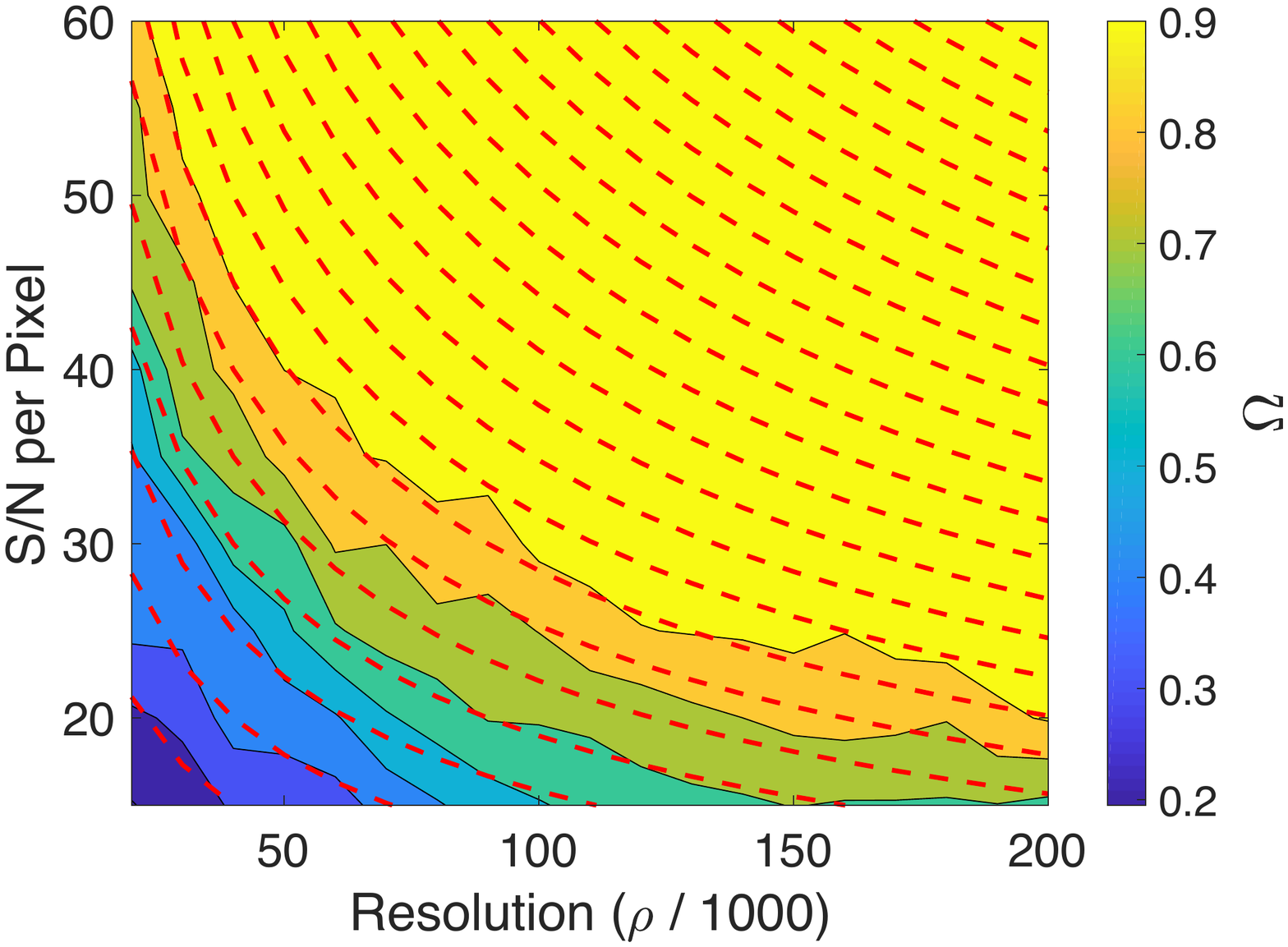}	  	
        (b)\includegraphics[trim = 0 150 0 120, clip, width = 0.45\textwidth]{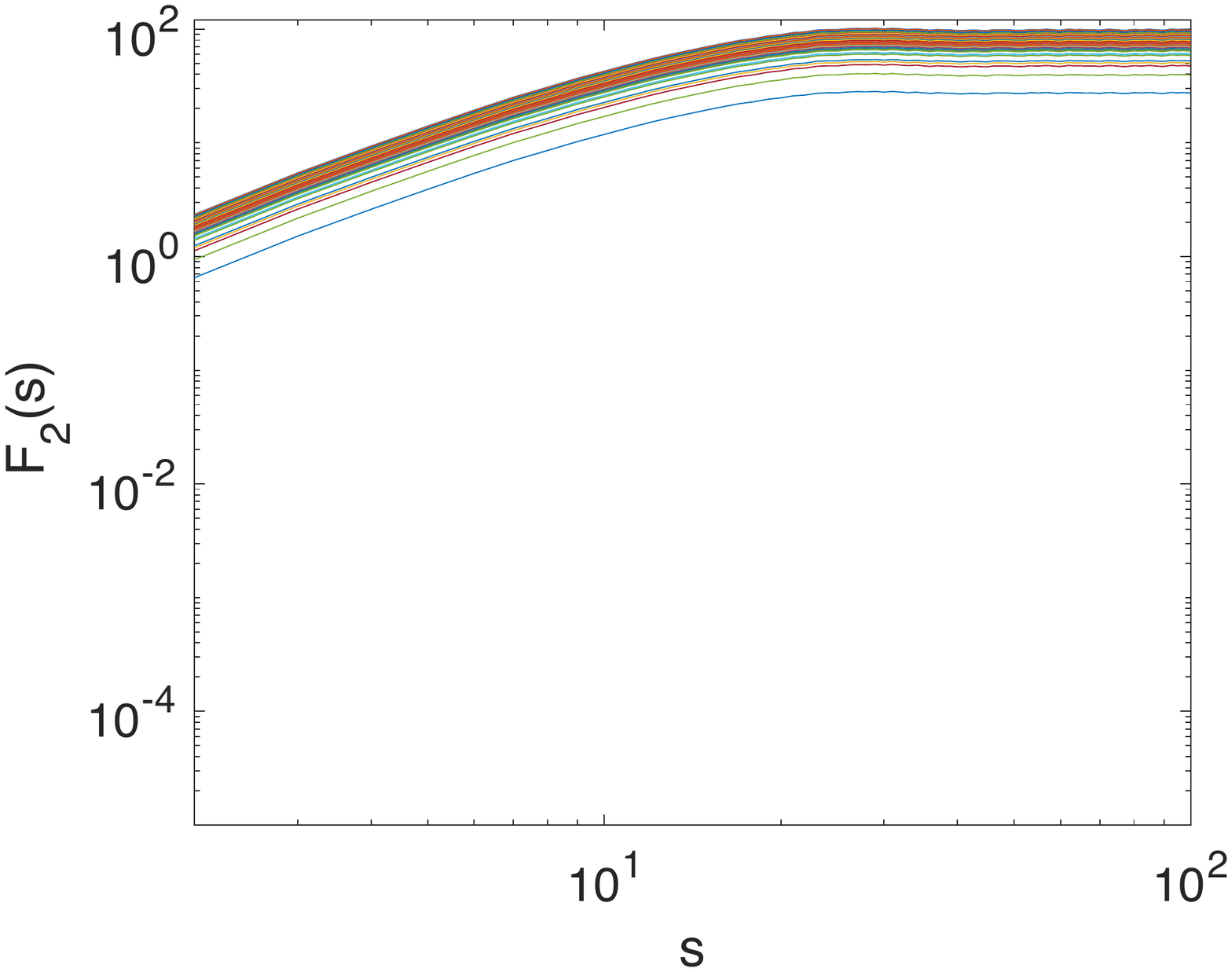}
        (c)\includegraphics[trim = 0 150 0 120, clip, width = 0.45\textwidth]{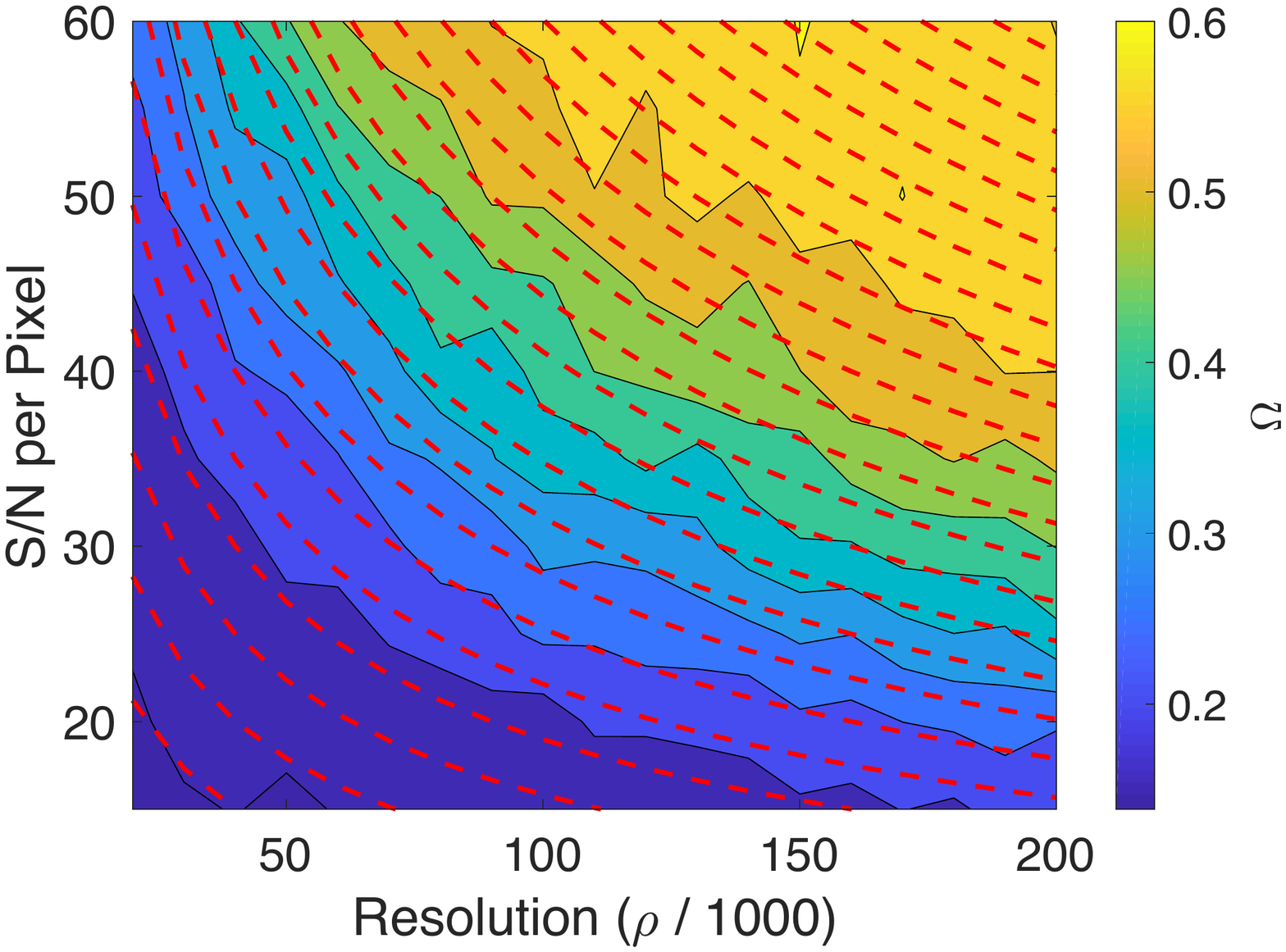}
        (d)\includegraphics[trim = 0 150 0 120, clip, width = 0.45\textwidth]{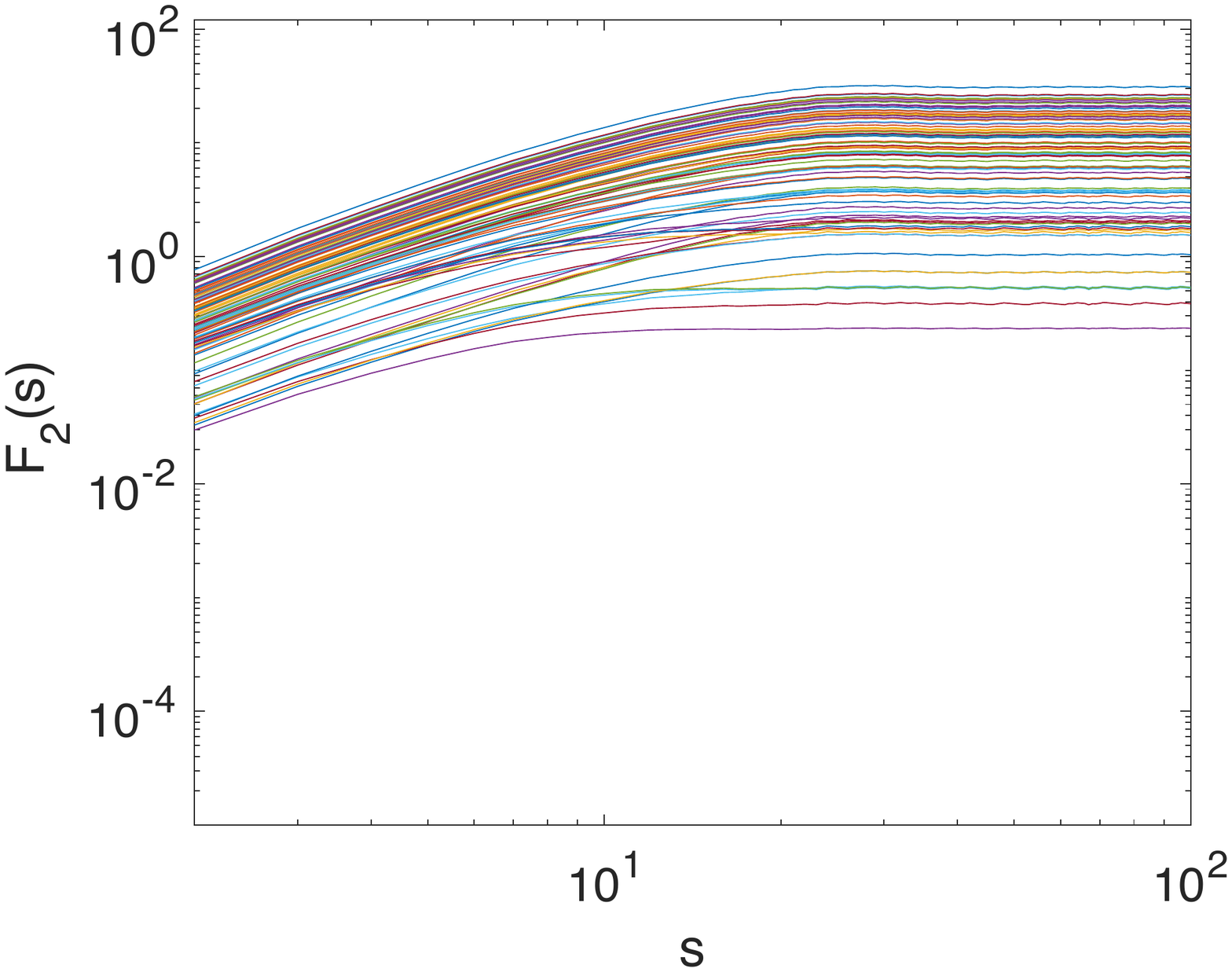}	  	
        (e)\includegraphics[trim = 0 150 0 120, clip, width = 0.45\textwidth]{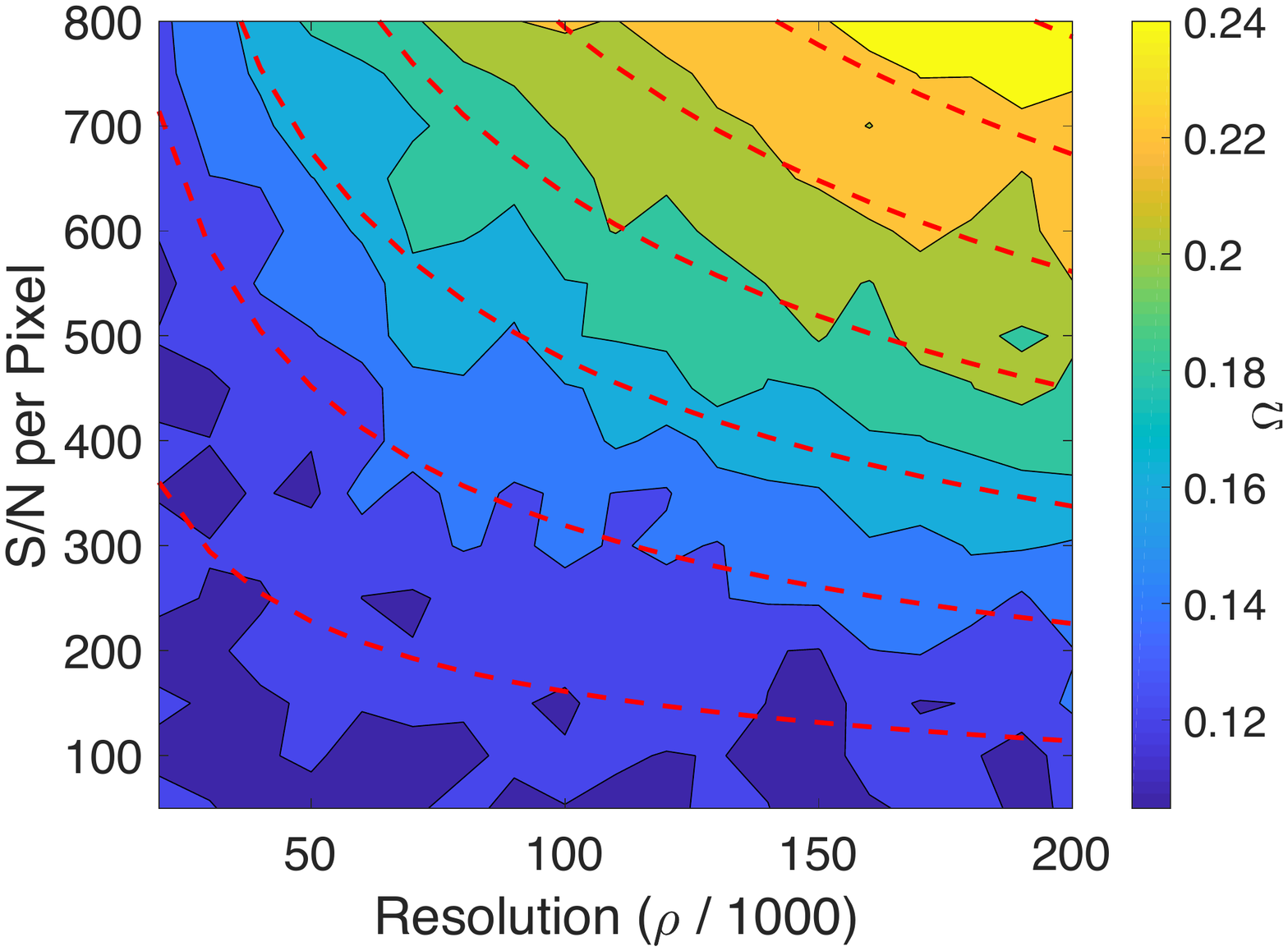}
        (f)\includegraphics[trim = 0 150 0 120, clip, width = 0.45\textwidth]{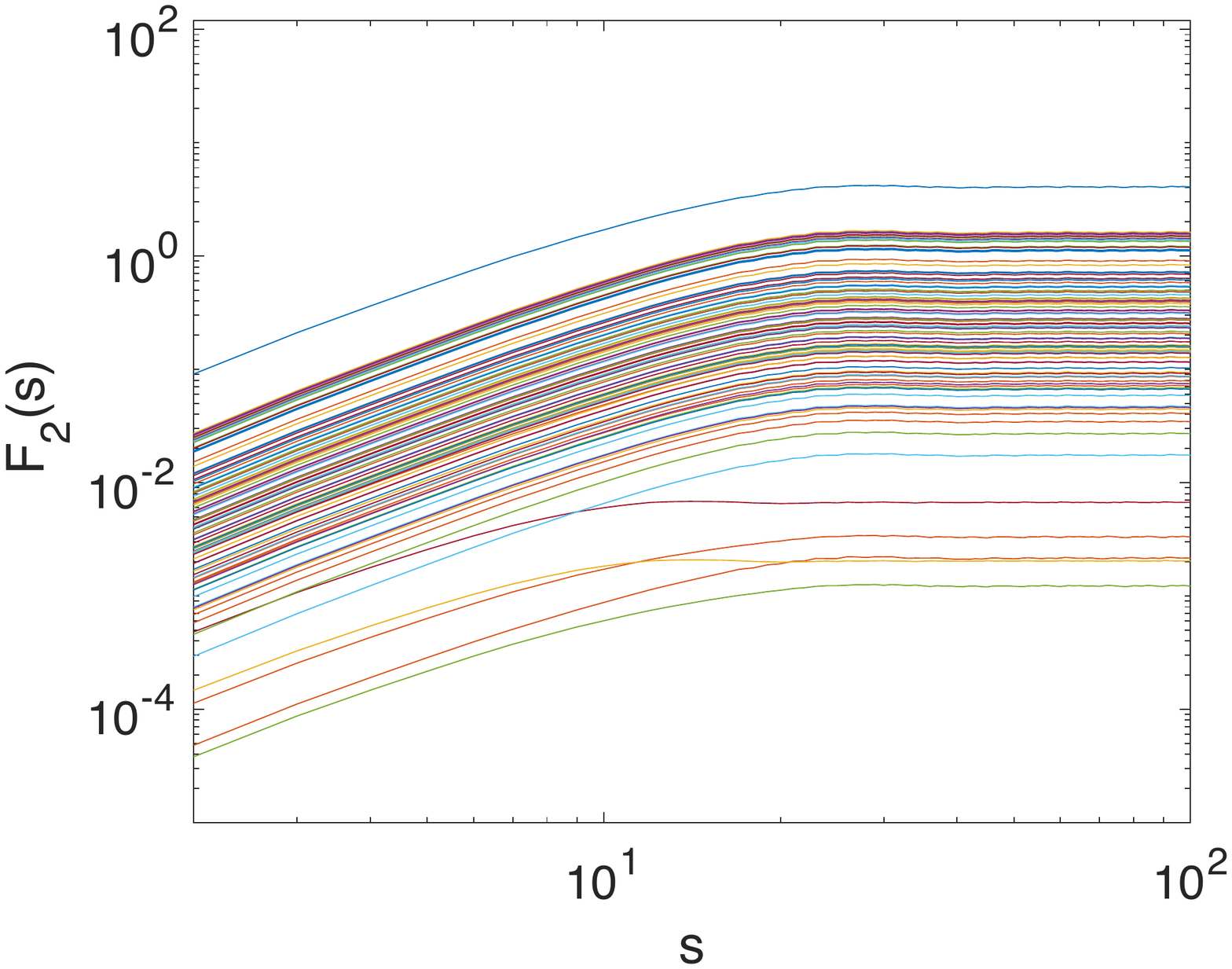}
    \caption{Contour plots for the ratio $\Omega$ from Eqn. \ref{eq:omega} and the corresponding fluctuation functions (no-noise and $\rho = 200,000$) for two stellar features (a-d) and a stellar companion (e-f). (a, b) A spot with 1\% coverage of the stellar disk; (c) A facula with 1\% coverage of the stellar disk; (e) Planet with a Radial Velocity of 10 ms$^{-1}$. Whereas PCA \citep{Davis:2017aa} requires a S/N $\sim 150$ (per resolution element or pixel) to detect a spot of this size, our method can detect it for a S/N $\sim 15$ (per pixel). This also shows how utilizing the statistical information of noise in the data rather than filtering it out refines the detection of such features.  In particular, the second moment of the fluctuation functions show that the several orders of magnitude decrease (in the fluctuations) of spots to faculae to planets corresponds to an increased sensitivity to noise, thereby requiring a higher S/N for detection.
Moreover, the inter-wavelength spread of these fluctuations as well as wavelength dependent fluctuations of faculae and planets can be used in differentiating between these features.
%The second moment of the fluctuation functions demonstrate a key property of their corresponding features, namely, the decrease by several  orders of magnitude of the fluctuations going from a spot to a facula to a planet denotes increased sensitivity to noise requiring a higher S/N for detection. 
%Another feature is the inter-wavelength spread of these fluctuations as well as wavelength dependent fluctuations in facula and planet which can be used in differentiating between these features. 
The red dashed lines in (a,c,e) are lines of equal photon flux. These closely parallel to the contour lines of $\Omega$, showing that MF-TW-DFA  captures information at low-resolutions, which has important implications for studying existing and future datasets.}
    \label{fig:CO_Noise}
\end{figure*}

\section{Data}

We use simulated data from the Spot Oscillation And Planet (SOAP) 2.0 tool \citep{Dumusque:2014aa}. This tool simulates a rotating star using a grid of cells filled with a clean solar spectrum from the National Solar Observatory. A spot or a facula is simulated by replacing the spectrum in one or more of the grid cells (depending on the size of the feature) by a spectrum of the spot or facula (with flux scaled in the spot spectrum according to the contrast ratio between a facula and the photosphere), which then follows the rotation of the star. The final spectrum is obtained by summing all the spectra in the grid. These spectra are \emph{noise-free} (S/N $\sim$ 1000) and have very high resolution (the spectra constitute $\sim$ 500,000 data-points and {the spectral resolution of a spectrograph refers to the precision with which it distinguishes wavelengths in the spectrum and is defined as $\rho = \lambda / \Delta \lambda$, where $\Delta \lambda$ is the smallest difference in all resolved wavelengths in a spectrum.  As resolution increases, so do the number of wavelengths resolved).} Thus we obtain three different types of spectra (spot, facula, planet) from this tool, where we can change the size of the feature or the planet induced radial velocity.  For simplicity,  the spot and facula are present on the equator of the star and the axis of rotation of the star and that of the planetary orbit (circular) are kept fixed at $90^\circ$.
%For this analysis, the spot and facula are present on the equator of the star while the axis of rotation of the star and that of the planetary orbit (circular) are kept fixed at $90^\circ$ for simplicity. 
We analyze three datasets with the following parameters:
\begin{itemize}
\item[1.] Spot: $A = 1\%$,
\item[2.] Facula: $A = 1\%$, and
\item[3.] Planet: $K$ = 10ms$^{-1}$,
\end{itemize}
where $A = \frac{R_{S/F}^2}{R_{Star}^2} \times 100\%$ is the percent area of the stellar disk covered by the feature, $K$ is the radial velocity, $R_{S/F}$ is the radius of the spot or facula and $R_{Star}$ is the radius of the stellar disk. These spectra have a period of 25 time units, which are then stacked 8 times to produce 200 spectra in time \citep{Agarwal:2017aa}.

In reality, we never have clean spectra and it has been a technological challenge in instrument design to build spectrographs with extremely high resolutions. Therefore, following the methods of \citet{Davis:2017aa}, we (a) decrease the resolution of these clean spectra and (b) add noise to them to model real observations. The resolution is decreased by convolving the original spectrum with a Gaussian having a full width at half maximum %(FWHM) 
equal to $ \lambda / \rho$, where $\lambda$ is a wavelength in the spectrum and $\rho$ is the desired resolution. Gaussian white noise per pixel is then added in the spectrum depending on the desired S/N. Steps (a) and (b) are performed for all six datasets, with the aim to obtain the lower limits of S/N and $\rho$ for such features to be detectable. For ease of computation, the final spectrum is then averaged to obtain a wavelength resolution of 1~\si{\angstrom}. We have also examined the original dataset to show that this averaging does not impact the results \citep{Agarwal:2017aa}.

\section{Methods}

\subsection{Multi-fractal Temporally Weighted Detrended Fluctuation Analysis}

The essence of a multifractal system is the exhibition of different self-similar structure on multiple different spatial or temporal (or both) scales \cite[e.g.,][]{Halsey:1986}.   In highly nonlinear coupled dynamical systems, the intrinsic dynamics and the extrinsic effects, such as noise, nearly always operate on multiple spatial and temporal scales, rather than exhibiting a single scaling. This underlies the use of a multifractal method. 

Many methods used to detect stellar features are influenced by the noise present in the data and therefore include a noise reduction strategy. 
To remove the bias induced by the noise removal strategy, or to define what qualifies as \emph{noise}, we instead use a method that utilizes the statistical information present in the noisy data to characterize a stellar feature. Given $m$ evenly spaced in time spectra, with each spectra spanning $L$ wavelengths, we construct $L$ time-series of length $m$ for each of the wavelengths. Here, each wavelength is analyzed separately, providing robustness to the results, as well as allowing us to distinguish between features that are present across all wavelengths versus those that are wavelength specific.

%To get rid of the bias induced by how the noise is removed, or what characterizes as \emph{noise}, we instead use a method that utilizes the statistical information present in the noisy data to characterize a stellar feature. Given $m$ evenly spaced in time spectra, with each spectra spanning $L$ wavelengths, we construct $L$ time-series of length $m$ for each of the wavelengths. Here, each wavelength is analyzed separately, providing robustness to the results, as well as allowing us to distinguish between features that are present across all wavelengths vs. those that are wavelength specific.

We analyze all $L$ time series using Multi-fractal Temporally Weighted Detrended Fluctuation Analysis (MF-TWDFA) \cite[See Appendix \ref{Sec:method},][and references therein]{Sahil:MF, Agarwal:2017aa}, which does not \textit{a priori} assume anything about the structure of the data, save for its multifractality.  Thus, we begin with an agnostic view about the system timescales and their corresponding dynamics. Moreover, rather than ``removing'' noise from the signal, we use it as a source of information on the dynamics of the system on all timescales present.  
%MF-TW-DFA  only assumes a multifractal structure of the system, so that we remain agnostic about the timescales and their corresponding dynamics the system might exhibit. 

The approach has four stages, which we describe in Appendix \ref{Sec:method}.  We produce a statistical measure called the fluctuation function, $F_q (s)$, each moment of which, $q$, is assessed on multiple time scales, $s$.  For intuition, one can think of the expectation value of $F_q (s)$ as the weighted sum of the auto-correlation function.  The dominant time scales in a system are the those where $F_q (s)$ versus $s$ changes slope and the individual slopes are associated with the statistical dynamics of a system.

\subsection{Principal Component Analysis}

%The essence of multifractals comes from the ability of the system to exhibit different dynamics on multiple scales. It is almost always the case that given a highly nonlinear coupled dynamical system, the intrinsic dynamics of the system as well as the extrinsic effects such as noise operate on multiple spatial and temporal scales, rather than exhibit a single scaling. This coupling on multiple scales inspires us to use a method well suited to analyze multifractal structures in the data.

Another method that is widely used to extract the dynamics of a system is Principal Component Analysis (PCA) \citep{Pearson:1901aa}, which has a history that predates the discovery of multifractals.  In PCA one transforms the underlying data into a set of linearly-independent orthogonal vectors, with their directions set by the directions of maximum variance. Given a set of observations, one computes the covariance matrix of the original data, $\Sigma$, upon which one performs singular value decomposition. The resulting diagonal matrix $D$, such that $\Sigma = UDV^T$,  gives the ordered set of eigenvalues denoting the amount of variance in the corresponding eigenvectors that are the columns of $U$. PCA has been very successful in analyzing data to, for example, to separate the dominant climate modes and long-term global temperature patterns on Earth \citep{Mann:1998aa, Mantua:2002aa, Moon:2018aa}, and to detect exoplanet and stellar features \citep{Thatte:2010aa, Poppenhaeger:2011aa, Davis:2017aa}. PCA has also been used widely as a noise reduction technique in which the complete system is projected onto the eigenvectors corresponding to a threshold (low) variability of the system. But this may also result in losing information about the system if the processes corresponding to ``noise'' actually represent the low variability dynamics of the system. Importantly, PCA also assumes that the system is governed by a Gaussian distribution and that the dynamics corresponding to each of the different modes is linearly independent. These assumptions are obviously not met {\em a priori} in a highly nonlinear system such as Earth's climate or the dynamics of a star-planet system.

\section{Discussion}

Figures \ref{fig:CO_Orig}~(a-c) show the crossover timescales from the fluctuation functions for the original data set with no noise for a spot, facula and planet, respectively. The main characteristics of  these plots are: (i) Almost all wavelengths show the same crossover timescale of 25 units, which is the rotational period of the star in the case of a spot and facula, and the orbital period of the star in the case of the planet, (ii) the spot has a clean timescale of 25 units  while the facula shows a wavelength specific pattern, since not all wavelengths are affected by it \citep{Davis:2017aa}, and (iii) the planet has some scatter around the timescale of 25 units. This provides a test bed for detecting such features as we degrade the simulation data.  We have also tested our method by performing continuum normalization of the spectra to account for changes in observational conditions, which we describe in Appendix \ref{Sec:normalized}. We show that normalizing the spectra only acts as a filter on the data, and thus does not change our analysis and results, as described below.
%in noisy data, which we can create by degradation.
%where we can match the above characteristics to the degraded datasets. 

To examine how the resolution and noise impacts these detections, we vary the resolution from 20,000 to 200,000 and vary the S/N from 15 to 60 in the case of the facula and the spot and S/N from 50 to 800 for the planet. We define a statistic $\Omega$,
\begin{equation}
\Omega = \frac{\text{Number of wavelengths with a timescale of 25 units}}{\text{Total Wavelengths}},
\label{eq:omega}
\end{equation}
which is the ratio of number of wavelengths with a timescale of approximately 25 units to the total number of wavelengths. 
Given some S/N and resolution of the underlying spectra, $\Omega$ is the probability of detecting a stellar feature or a planet in the data. Thus, depending on the goals of a particular application, one can define a threshold probability for detecting a feature, for example defining $\Omega_{Th} = 0.5$ implies that any feature with $\Omega \ge 0.5$ would correspond to a robust detection. Clearly, any value of $\Omega_{Th}$ has an important trade-off in terms of balancing the false positives and false negatives.

%{\color{blue} 
To compare MF-TW-DFA with PCA, we construct a metric $\mathcal{F}_{PC}$, in a manner similar to $N_{PC}$ of \citet{Davis:2017aa}. While $N_{PC}$ gives approximately the number of principal components in the data that can be used to extract useful information about the system, it does not quantify that information. $N_{PC}$ is a whole number, computed by rounding up the average correlations between {the scores (defined as the projection of the data matrix onto its eigenvectors)} of the noisy data and the ideal data. Therefore, we have constructed a metric $\mathcal{F}_{PC}$ that measures the fraction of the total variability in these aforementioned principal components, and thus the amount of information in the data as measured by PCA.  Hence, $\mathcal{F}_{PC}$ provides a more refined description of the variability in the data than does $N_{PC}$, which loses such detail due to the rounding process.  
%Due to rounding up in $N_{PC}$, finer detail is lost and thus $\mathcal{F}_{PC}$ provides a finer description of the variability in the data. 
Since, $\Omega$ is a measure of probability in wavelength space, it should also be thought of as the information content in the data as measured by MF-TW-DFA. 
Irrespective of the value of $N_{PC}$, both $\mathcal{F}_{PC}$ and $\Omega$ are defined on a scale of [0,1] thereby allowing us to perform a direct comparison between the two methods. 

Similar to $N_{PC}$, we first compute the correlation between the first 10 scores of the noisy data (given S/N and resolution) and the ideal data (original S/N and resolution).  We then compute the p-value and choose those eigenvectors/eigenvalues for which the correlation has p-value $< 0.001$ \citep{Davis:2017aa}.  The definition of $\mathcal{F}_{PC}$ is 
\begin{equation}
\mathcal{F}_{PC} \equiv \sum_{i = 1}^{10} \sum_{j = 1}^{50} \mathcal{I}_{R_{ij}} \frac{f_{ij}}{50},
\label{eq:fpc}
\end{equation}
where $f_{ij}$ is the fraction of total variability given by the $i_{th}$ eigenvector of the $j_{th}$ ensemble member, $R_{ij}$ is the absolute correlation coefficient between the $i_{th}$ score and the $j_{th}$ ensemble member of the noisy data and the $i_{th}$ score of the ideal data, and 
\begin{equation}
\mathcal{I}_{R_{ij}} = \begin{cases}
1 & \text{ if the p-value for } R_{ij} < 0.001 \\ 
0 & \text{ otherwise} .
\end{cases}
\label{eq:Ipc}
\end{equation}

\begin{figure*}[htbp!]
        (a)\includegraphics[trim = 0 150 0 120, clip, width = 0.45\textwidth]{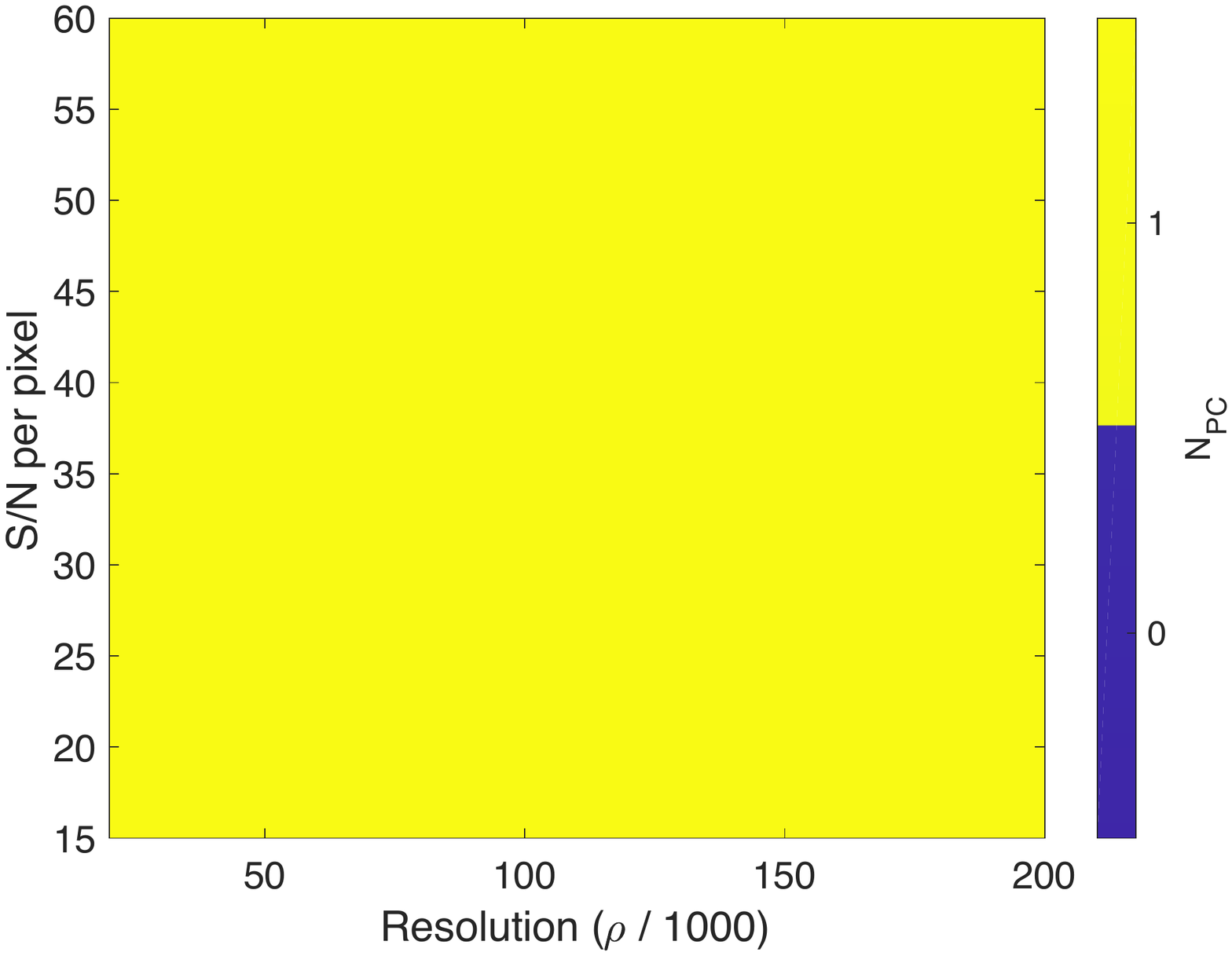}	  	
        (b)\includegraphics[trim = 0 150 0 120, clip, width = 0.45\textwidth]{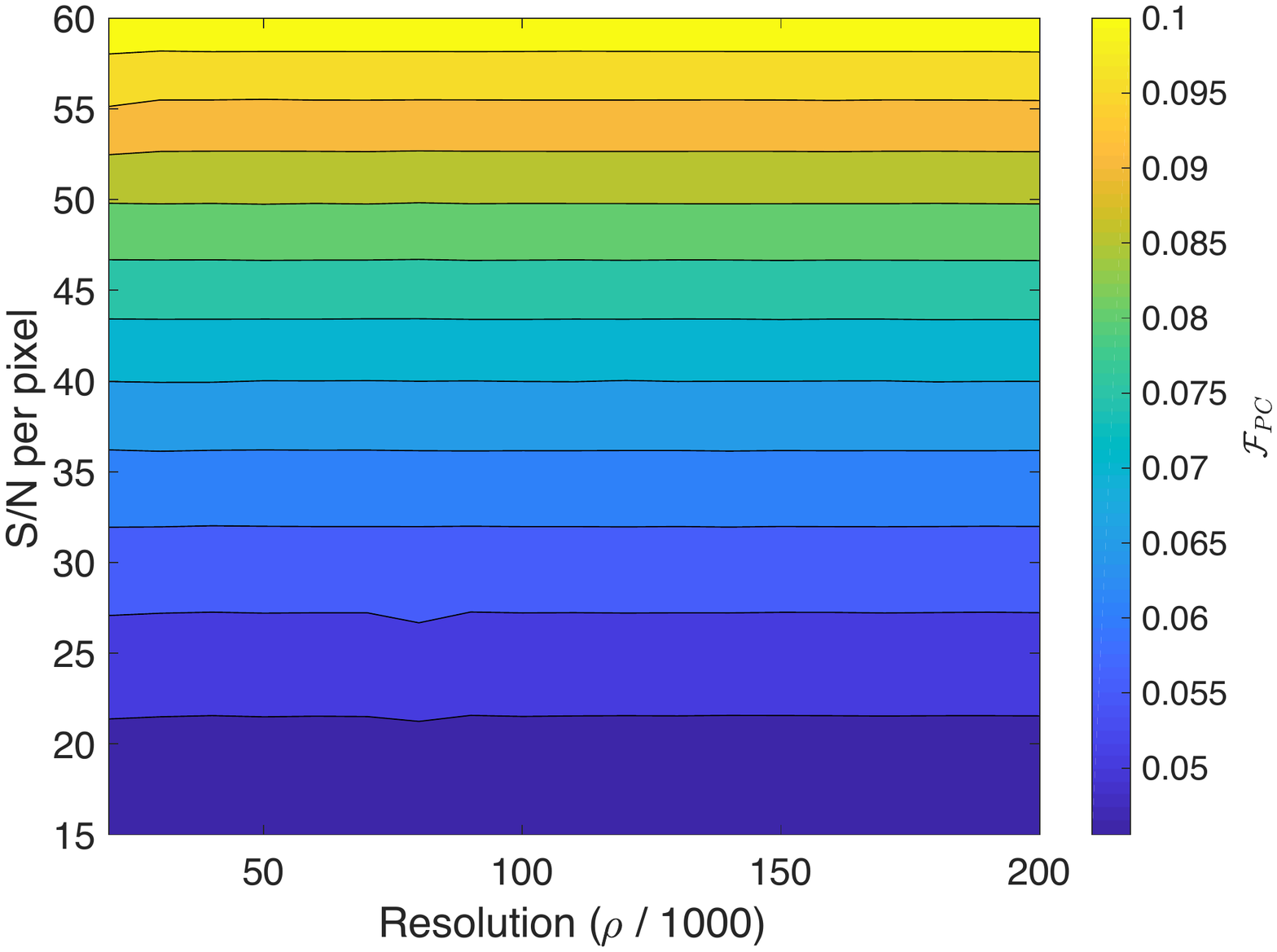}
        (c)\includegraphics[trim = 00 150 0 120, clip, width = 0.45\textwidth]{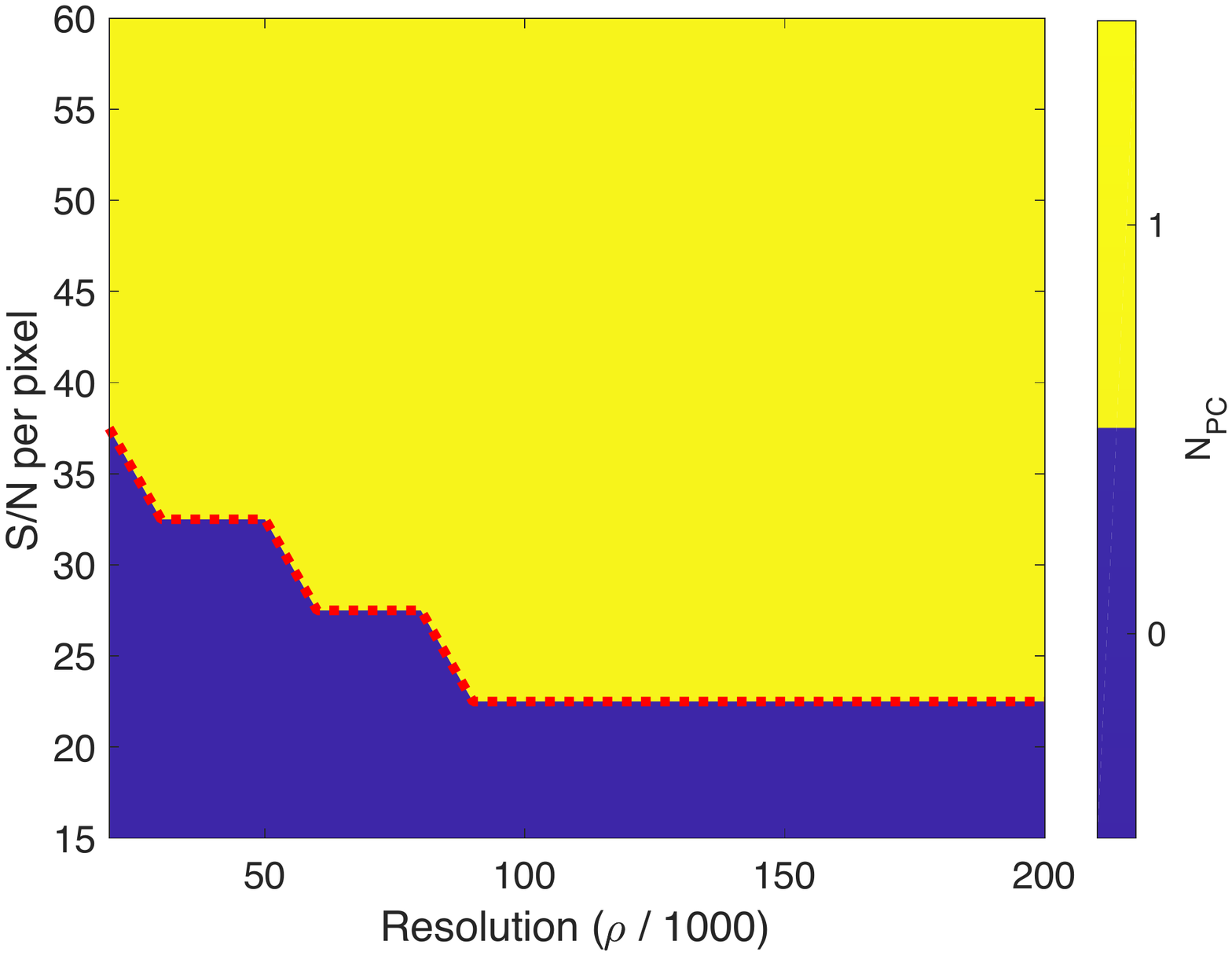}
        (d)\includegraphics[trim = 0 150 0 120, clip, width = 0.45\textwidth]{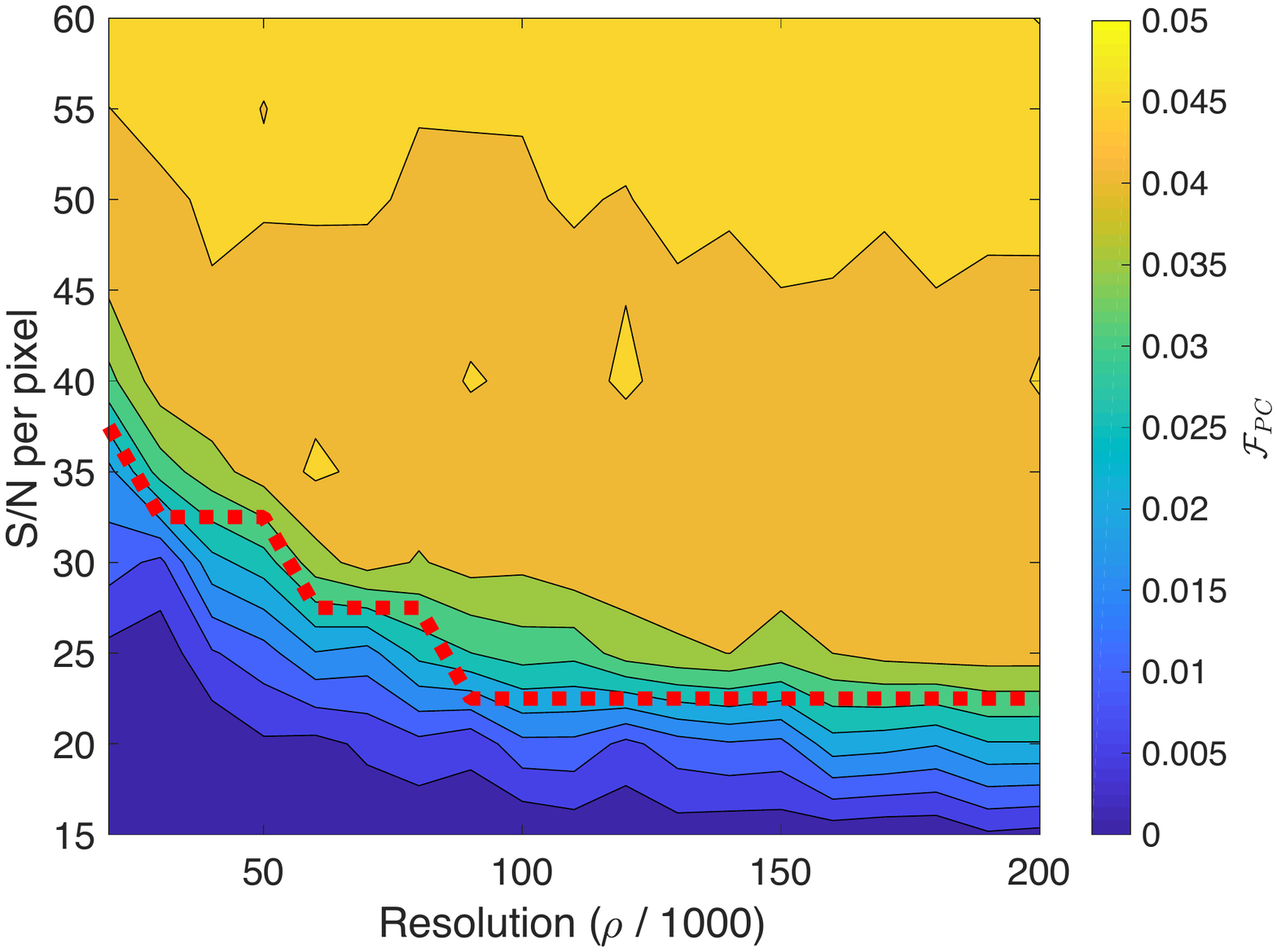}	  	
        (e)\includegraphics[trim = 0 150 0 120, clip, width = 0.45\textwidth]{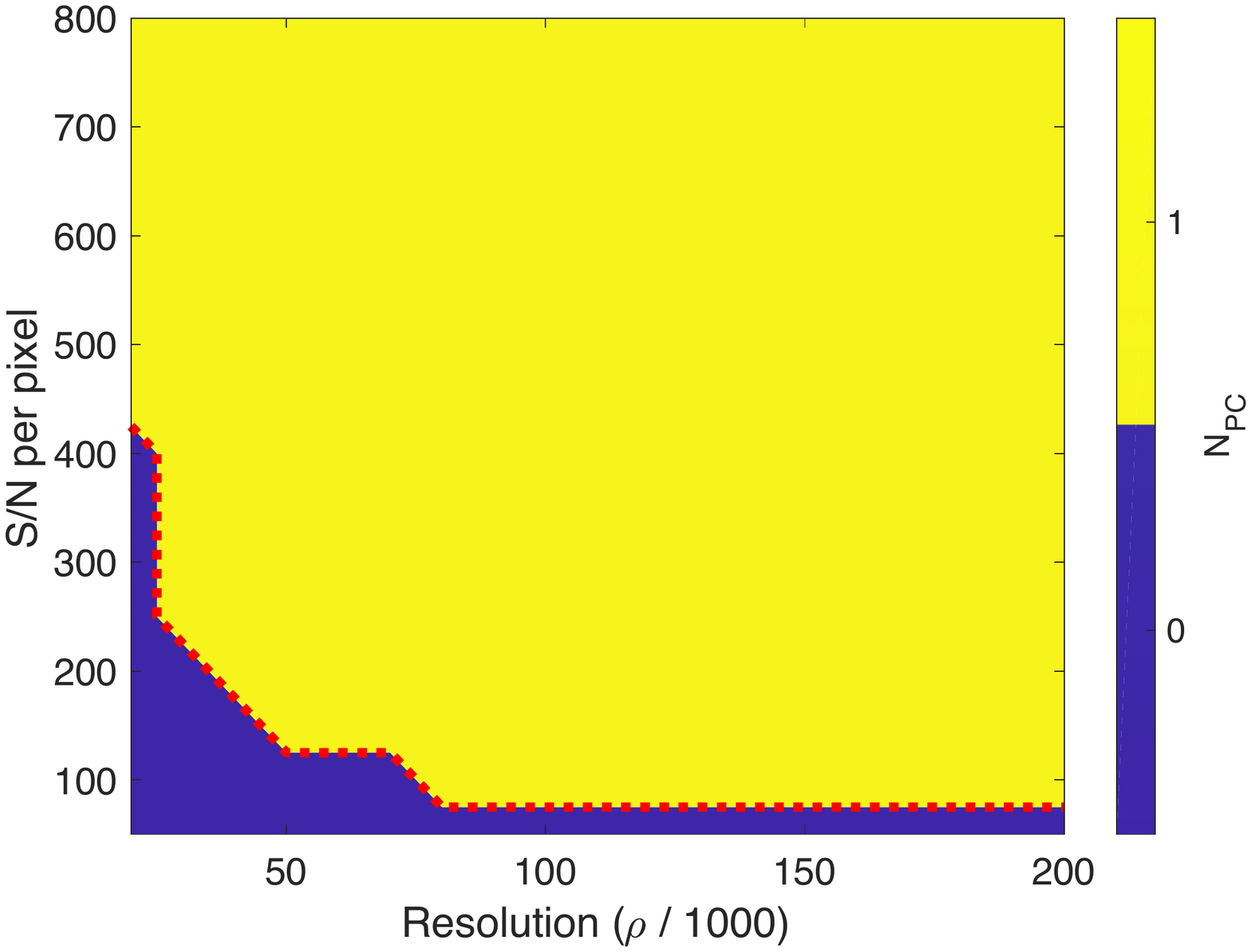}
        (f)\includegraphics[trim = 0 150 0 120, clip, width = 0.45\textwidth]{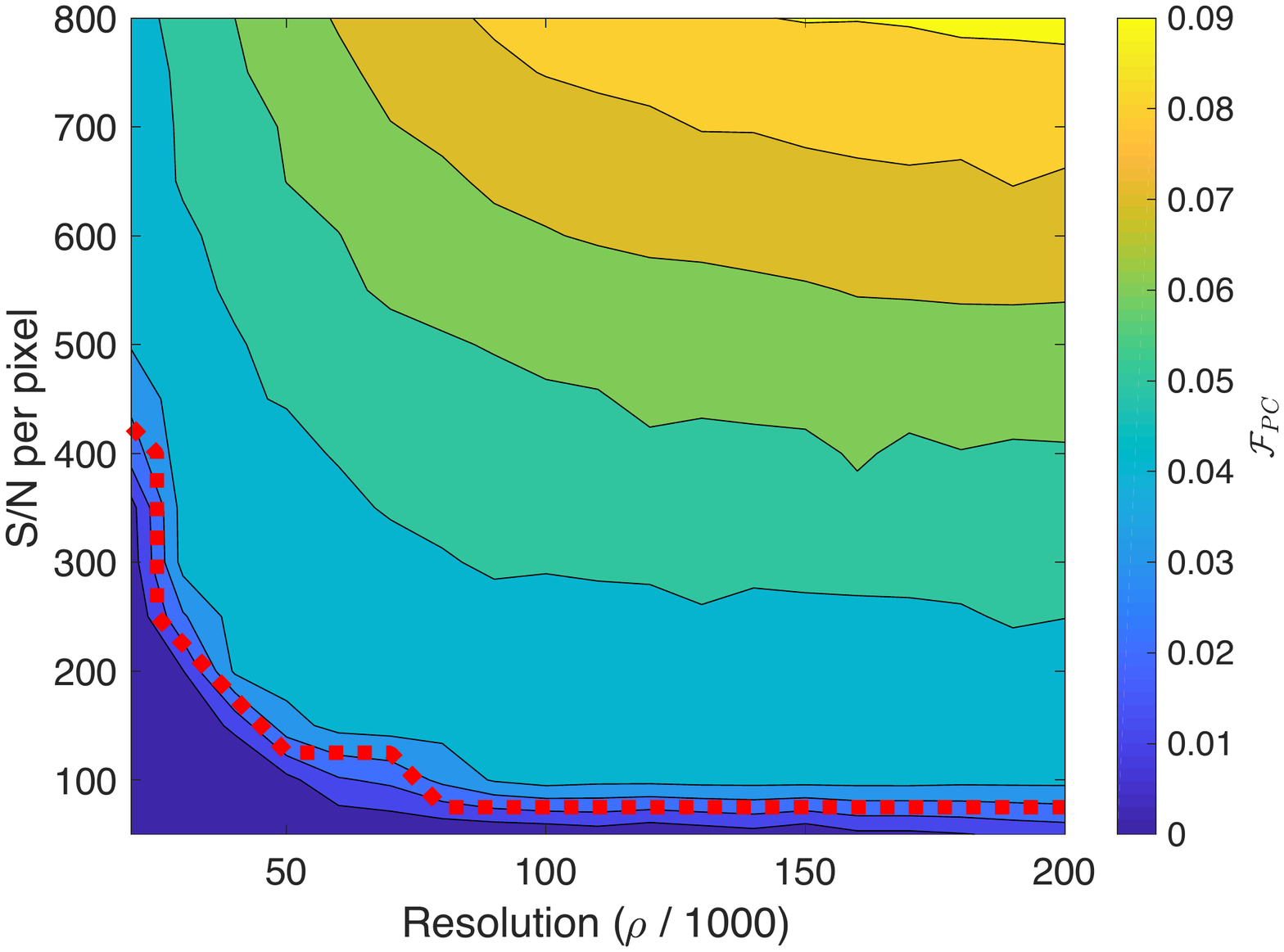}
    \caption{Contour plots for $N_{PC}$ computed as in \citet{Davis:2017aa} and $\mathcal{F}_{PC}$ from Equation \ref{eq:fpc} for a spot (a,b), a facula (c,d) and a planet (e,f). Comparing the values of $\mathcal{F}_{PC}$ here with $\Omega$ in Figure \ref{fig:CO_Noise} shows that MF-TW-DFA provides information on a scale that is at least an order of magnitude finer than that provided by PCA and thus has a clear advantage in capturing stellar features and nonlinear dynamics. The dotted red line in the $\mathcal{F}_{PC}$ plots in (d) and (f) is the contour line from the corresponding $N_{PC}$ plots in (c) and (e). Thus, while $N_{PC}$ gives a whole number, $\mathcal{F}_{PC}$ provides a fine structure in the same data range. }
    \label{fig:PCA}
\end{figure*}

For the same parameters as in Figure \ref{fig:CO_Noise} we show $N_{PC}$ and the corresponding $\mathcal{F}_{PC}$ for a spot, a facula and a planet in Figures \ref{fig:PCA} (a-f).
%Figures \ref{fig:PCA} (a-f) show $N_{PC}$ and the corresponding $\mathcal{F}_{PC}$ for a spot, a facula and a planet, for the same parameters as in Figure \ref{fig:CO_Noise}. 
Both  $\Omega$ and $\mathcal{F}_{PC}$ provide high resolution measures of the amount of information in the noisy data.  
Clearly, at all S/N values and resolutions, MF-TW-DFA performs at least an order of magnitude better than PCA.   The superior performance MF-TW-DFA is principally due to the fact that the method (a) is not restricted to capturing linear dependencies in the data as is PCA, and (b) uses `noise' as a source of information.  

\vspace{-0.25 cm}
\subsection{S/N versus Resolution}
\vspace{-0.25 cm}

Figures \ref{fig:CO_Noise}~(a,c,e) show contour plots of $\Omega$ as a function of the resolution $\rho$ and S/N. The robustness of our method is clear: 
%which making no a priori assumptions about  the temporal structure of the data under study, is clear: 
We can detect a spot with S/N as low as 15 while for S/N $\ge 35$, the results are independent of resolution.  Because a facula exhibits noise behavior that differs from a spot, detecting it requires a higher S/N value, but still as low as 45.  It is useful to compare this with Fig. 7 of \citet{Davis:2017aa} (their y-axis is S/N per resolution element), which is a PCA based study.  

%We compare this with Fig. 7 of \citep{Davis:2017aa} (their y-axis is S/N per resolution element). MF-TW-DFA does not a-priori assume any temporal structure of the data under study, and thus the statistical information extracted from this methodology is much more robust. This is seen from Fig. \ref{fig:CO_Noise}, where our method can detect a spot with S/N going as low as 15 while for S/N $\ge 35$, the resolution does not even matter. Because a facula shows different noise behavior as compared to a spot, it's detection requires a higher S/N value (as low as 45).

By comparing the lines of equal photon flux (S/N $\propto$ $1/\rho$) with the contour lines in their Fig. 7, \citet{Davis:2017aa} demonstrated that in order to identify stellar signals in spectral data, having high resolution is more important than having high S/N.  In contrast, we see that the red dashed lines in Figure \ref{fig:CO_Noise} (a,c,e) are the lines of equal photon flux and run parallel to the contour lines of $\Omega$, showing that with MF-TW-DFA low-resolution spectra can be successfully utilized to capture and study these stellar phenomena. 
%our method shows that this is not the case and low-resolution spectra can be successfully utilized to capture and study these stellar phenomena. The red dashed lines in Figure \ref{fig:CO_Noise} (a,c,e) are the lines of equal photon flux, and run parallel to the contour lines of $\Omega$, and thus MF-TW-DFA does not need high-resolution data to capture stellar phenomenon. 
This has significant implications for the analysis of both existing datasets and incipient missions, for which we can utilize the multifractal methodology described here to extract high quality information from what may have previously been viewed as noisy and/or low-resolution data.

\subsection{Fluctuation Strength}

Figures \ref{fig:CO_Noise}~(b,d,f) show the second moment of the fluctuation functions from Equation (\ref{eq:power}) for all wavelengths. The main characteristics of these plots are: (i) Ostensibly all wavelengths are parallel to each other, with only a minute number of wavelengths deviating from the bulk (facula and planet) (see also Fig. \ref{fig:CO_Orig}), (ii) a substantial decrease in the fluctuation strength as we transition from spots to faculae to planets.  The implication of this decrease in fluctuation strength is an increase in the sensitivity of fluctuation functions to noise, thereby requiring a higher S/N to robustly detect a feature. 
%going from the spot to the facula to the planet (Figs. \ref{fig:CO_Noise}~(b,d,f)) we see a substantial decrease in the fluctuation strength on each transition. The implications of this decrease in fluctuation strength is that the lower the fluctuation strength, the more sensitive the fluctuation function is to the role of noise, thereby requiring a higher S/N to robustly detect a feature. 
%Therefore the we see that as we move from spot to facula to planet, the required S/N for a robust detection increases.

\subsection{Differentiating spots, faculae and planets}

Figures \ref{fig:CO_Orig} and \ref{fig:CO_Noise}~(b,d,f) exhibit the essential property of the individual phenomenon discussed above that differentiates spots, faculae and planets. Namely, the abruptness and clarity of the crossover timescales (Fig. \ref{fig:CO_Orig}) and the strength of the fluctuations (Figs. \ref{fig:CO_Noise}~(b,d,f)). 
In the SOAP2.0 code, a simulated facula is generated using a slight alteration of the spot spectrum, as compared to the intrinsic line-by-line variability between a sunspot spectra and the photosphere.  Using PCA \citet{Davis:2017aa} found that the principal components of a spot are quite similar to those of a facula.  Here we have shown that MF-TW-DFA is very efficient at detecting the small fluctuations for the facula (Fig. \ref{fig:CO_Orig}) and thus is able to differentiate between a spot and a facula.  The fluctuation functions also differ with respect to the inter-wavelength spread, which is much smaller for the spot than for the planet.  This spot-to-planet increase in inter-wavelength spread is an important quantitative detection diagnostic intrinsic to MF-TW-DFA.  Indeed, the spectral line distinction between spots, plages, faculae and the presence of an exoplanet \citep{Thompson:2017vw, Wise:2018vn, Dumusque:2018ws} underlie the effectiveness of our detection framework.  This suite of 
characteristics can therefore be utilized to increase the robustness and fidelity of multi-feature detections in observations.

%denoting a much more uniform effect of a spot on all the wavelengths, whereas the planet exhibits a much larger spread in the fluctuations. 
%The different spectral lines are affected differently by spots,  plages, faculae and the presence of an exoplanet \cite{Thompson:2017vw, Wise:2018vn, Dumusque:2018ws} and MF-TW-DFA is effective in detecting these changes. These characteristics can therefore be utilized to differentiate between these phenomenon and in turn increase the robustness and fidelity of detections of these features in observations.

%Namely, the cleanliness of the crossover timescales (Figure \ref{fig:CO_Orig}) and the strength of the fluctuations (Figure \ref{fig:CO_Noise}(b,d,f)). The fluctuation functions also differ with respect to the spread amongst each other with being tightly packed for the spot signifying a much more uniform effect of a spot on all the wavelengths vs a planet that exhibits the maximum spread in the fluctuations. These characteristics can therefore be utilized in order to differentiate between these phenomenon and in turn increase the robustness and fidelity of detections of these features in observations.

%These results showcase the importance of using noise as a source of information rather than something to be filtered out.  

\section{Conclusion}

We have used simulation data of a stellar spectrum from the SOAP 2.0 tool in the presence of a spot, a facula or a planet to demonstrate the fidelity of a multifractal methodology for detecting exoplanets.   The motivation is to provide a framework that naturally deals with unavoidable noise in observational systems.  The approach is unique conceptually in that it uses noise as a source of information rather than treating it as something to be filtered out.  By using controlled simulation data we systematically varied both the resolution and the signal-to-noise (S/N) ratio to determine a lower limit on the resolution and S/N required to robustly detect features using our multifractal method.  For any resolution above 20,000, we found that a spot and a facula with a 1\% coverage of the stellar disk can be robustly detected for a S/N (per pixel) of 35 and 60 respectively, whereas a planet with a radial velocity of 10 ms$^{-1}$ can be detected for a S/N (per pixel) of 600.  These results are compared to the standard PCA method, which requires a S/N 100 times larger than our multifractal method to detect the same information.  A key aspect of these results, of relevance to considering which methodologies are most appropriate for observational data, is whether they have fidelity sufficient to discern stellar features themselves from planets.  Finally, the method allows for a systematic examination both of intrinsic stellar dynamical processes and practical noise sources, such as telluric or instrumental noise, to help refine observational schema.  

\acknowledgements

SA and JSW thank Debra Fischer for a series of extremely helpful conversations and they acknowledge NASA Grant NNH13ZDA001N-CRYO for support. JSW acknowledges Swedish Research Council grant no. 638 -2013-9243, and a Royal Society Wolfson Research Merit Award for support.

%\bibliography{ExoRefs}

\appendix

\section{Multi-Fractal Temporally Weighted Detrended Fluctuation Analysis}\label{Sec:method}

For a given set of evenly spaced spectra with $L$ wavelengths, we construct $L$ time series and analyze each time series independently. We analyze the time-series for each wavelength using Multi-Fractal Temporally Weighted Detrended Fluctuation Analysis (MF-TW-DFA) \citep[and references therein]{Sahil:MF}. This method has previously been used to study the temporal structure of Arctic sea ice extent \citep{Sahil:MF} and velocity fields \citep{Agarwal:2017bb}, exoplanet detection without the use of model fitting \citep{Agarwal:2017aa}, and multi-decadal global climate dynamics modes on Earth \citep{Moon:2018aa}.
It involves four steps, described here:
\begin{enumerate}

\item Given the original time series $X_i$, construct a non-stationary {\em profile} $Y(i)$ as, 
\begin{equation}
Y(i )\equiv \sum_{k=1}^{i} \left(X_k - \overline{X}~ \right), \qquad \text{where}\qquad  i = 1, ... , N.  
\label{eq:profile}
\end{equation}
The profile is the cumulative sum of the time series and $\overline{X}$ is the average of the time series $X_i$.

\item This non-stationary profile is divided into $N_s = \text{int}(N/s)$ non-overlapping segments of equal length $s$, where $s$ is an integer and varies in the interval $1<s\le N/2$.  Each value of $s$ represents a time scale $s \times \Delta t$, where $\Delta t$ is the temporal resolution of the time series.
The time series has a length that is rarely an exact multiple of $s$, which is handled by repeating the procedure from the end of the profile and returning to the beginning, thereby creating $2 N_s$ segments.  

\item A point by point approximation $\hat{y}_{\nu}(i)$ of the profile is made using a moving window, smaller than $s$ and weighted by separation between the points $j$ to the point $i$ in the time series such that $\vert i - j \vert \le s$.  Note that in regular MF-DFA, as opposed to MF-TW-DFA, $n$th order polynomials $y_{\nu}(i)$ are used to approximate $Y(i)$ {\em within} a fixed window, without reference to points in the profile outside that window.
%\cite{MF_Footnote}.%\footnote{In regular MF-DFA, rather than using temporal-weighting,  $n$th order polynomials $y_{\nu}(i)$ are used to approximate $Y(i)$ {\em within} a fixed window, without reference to points in the profile outside that window.} 
A larger (or smaller) weight $w_{ij}$ is given to $\hat{y}_{\nu}(i)$ according to whether $\vert i - j \vert $ is small (large) \citep[][]{Sahil:MF}. This approximated profile is then used to compute the variance spanning up ($\nu = 1,...,N_s$) and down ($\nu = N_s + 1,...,2 N_s$) the profile as
\begin{eqnarray}
\text{Var}(\nu, s) & \equiv  \frac{1}{s}  \sum_{i=1}^{s}& \{ Y([\nu - 1]s + i) \nonumber \\
					&& - {\hat{y}}([\nu-1]s +i) \}^2 \nonumber \\
&&\text{for $\nu = 1,...,N_s$ ~~~~and} \nonumber \\
%\nonumber \\
\text{Var}(\nu, s) & \equiv  \frac{1}{s}  \sum_{i=1}^{s}& \{ Y(N-[\nu - N_s]s + i) \nonumber \\
					&& - {\hat{y}}(N-[\nu-N_s]s +i)\}^2 \nonumber \\
&&\text{for $\nu =  N_s + 1,...,2 N_s$.}
\label{eq:varTW}
\end{eqnarray}

\item Finally, a {\em generalized fluctuation function} is obtained and written as
\begin{equation}
F_q (s) \equiv \left[ \frac{1}{2 N_s} \sum_{\nu=1}^{2 N_s} \{ \text{Var}(\nu, s)\}^{q/2} \right]^{1/q}.
\label{eq:fluct}
\end{equation}
\end{enumerate}

We study the behavior of $F_q(s)$ with respect to the timescale $s$. In particular, the generalized fluctuation function scales with the timescale $s$ for a given moment $q$,  with the scaling exponents given by the generalized Hurst exponents $h(q)$,
\begin{equation}
F_q (s) \propto s^{h(q)} .  
\label{eq:power}
\end{equation}

These Hurst exponents convey the statistical information in the data at some timescale. If $h(q)$ is independent of $q$ the data is said to be a mono-fractal. The second moment of the fluctuation function is generally studied more than others, since it can be related to more common power spectrum. For power spectrum $S(f) \propto f^{-\beta}$, where $f$ is the frequency and $\beta$ is its decay exponent, $h(2) = (1 + \beta)/2$ \citep{Ding}. Therefore, since for white noise $\beta = 0$, it gives $h(2) = 1/2$. Similarly for pink noise $\beta=1$, which gives $h(2) = 1$, for red noise $\beta = 2$, giving $h(2) = 3/2$, and so forth. This allows us to characterize the data using noise characteristics on multiple timescales, where the dominant timescales in the data set are the points where the fluctuation function $\log_{10}F_2(s)$ changes slope with respect to $\log_{10}s$. We apply this method to each wavelength in the spectrum, giving us the dominant timescales of the complete spectrum. 

%\textcolor{red}{Above. (a) No mention here at all of the spectra. (b) Fix A2.}\\

%\textcolor{red}{{\bf Bibliography}.  (a) Full Personal names are not used in Ref. list. (b) Something like Ref. 5 and 6 should have the first 3 authors then et al., for AJ/ApJ. (c) If AJ/ApJ then most of the journal titles must change.}
%{\color{blue}

\begin{figure*}[htbp!]
%\begin{figure}[htbp!]
        (a)\includegraphics[trim = 0 150 0 120, clip, width = 0.45\textwidth]{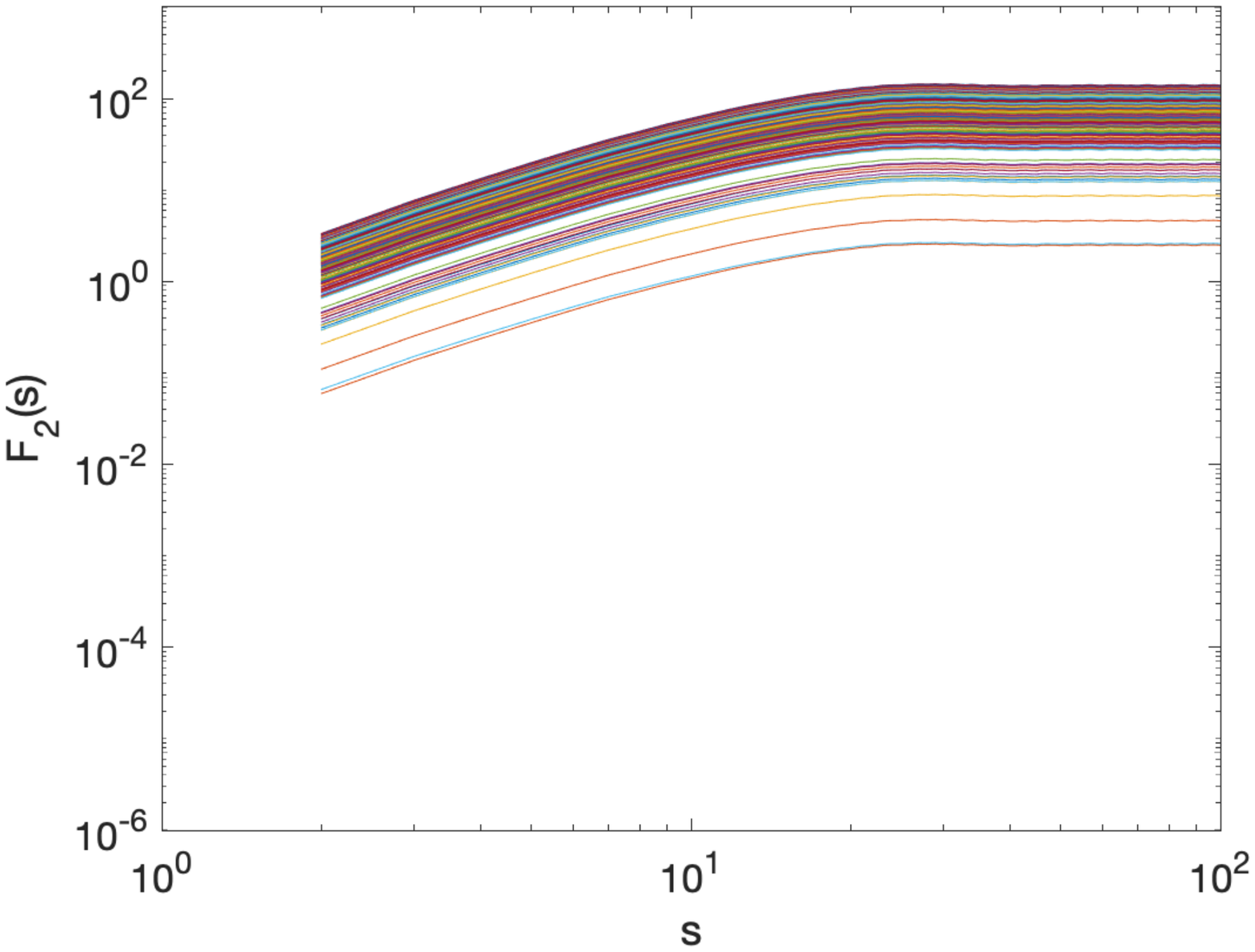}	  	
        (b)\includegraphics[trim = 0 150 0 120, clip, width = 0.45\textwidth]{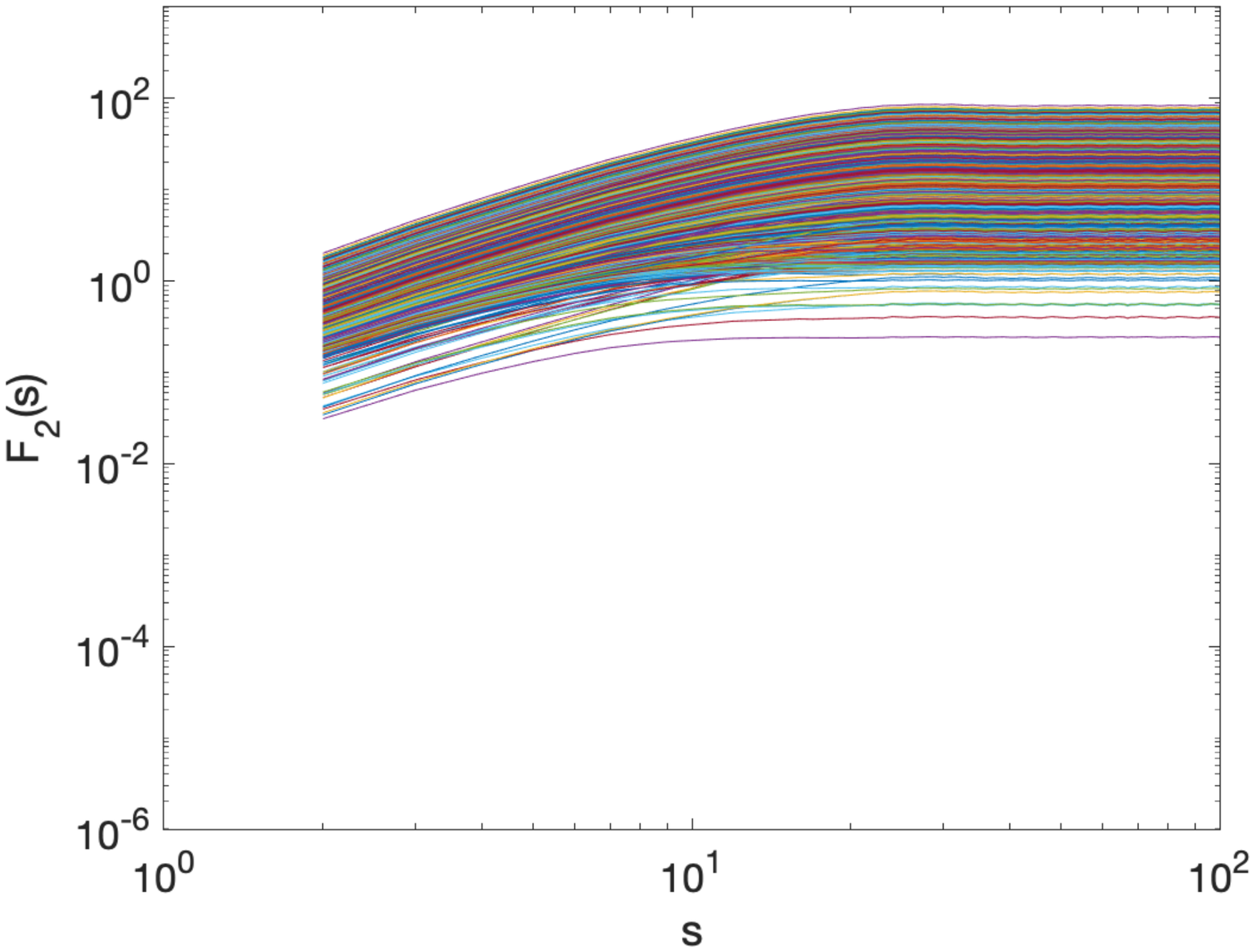}
        (c)\includegraphics[trim = 0 150 0 120, clip, width = 0.45\textwidth]{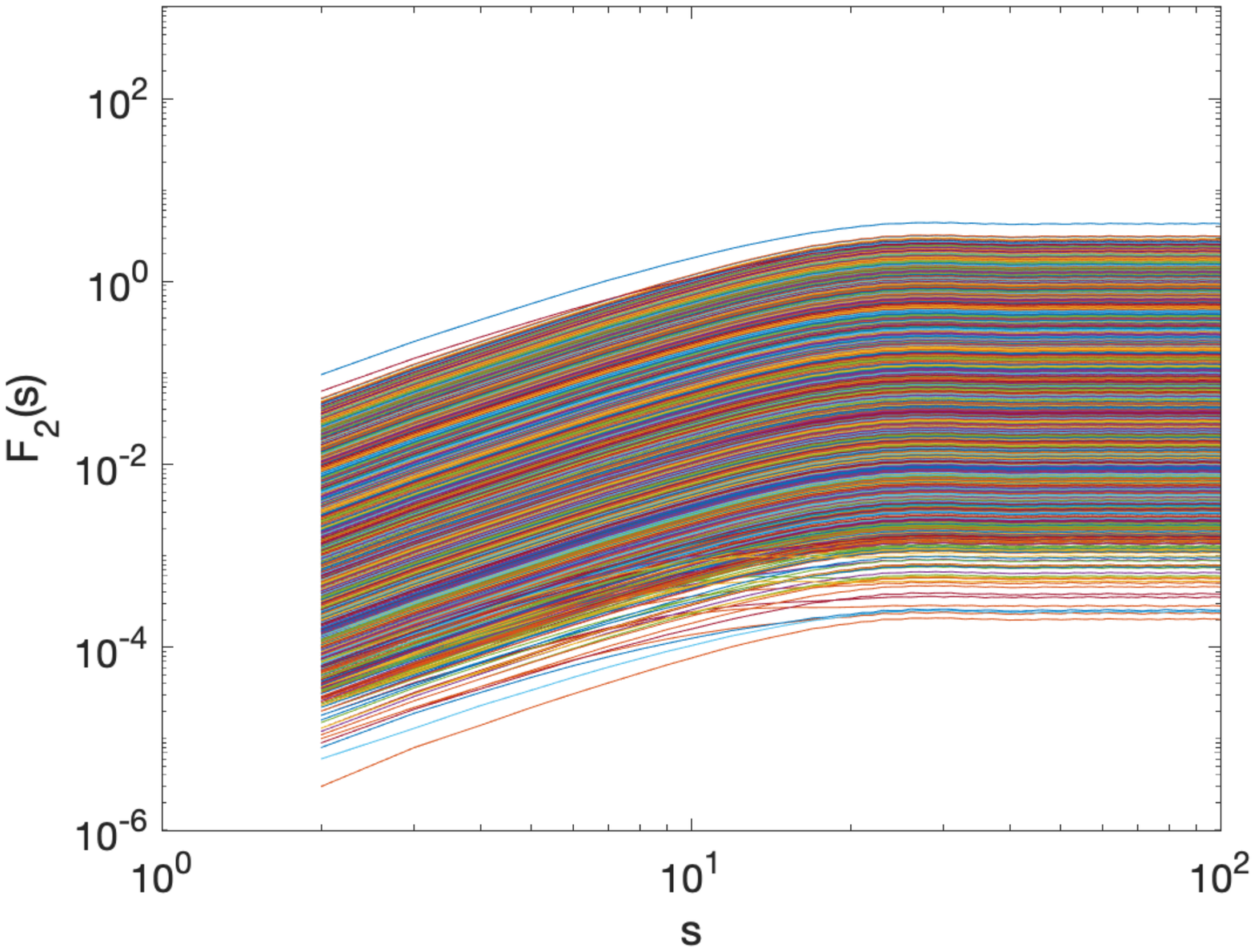}
    \caption{The fluctuation functions (no-noise and $\rho = 200,000$) for two stellar features (a, b) and a stellar companion (c). (a) A spot with 1\% coverage of the stellar disk; (b) A facula with 1\% coverage of the stellar disk; (c) Planet with a Radial Velocity of 10 ms$^{-1}$. Continuum normalization was performed on the data to account for changes in observational conditions. The normalization above was performed as the final step, after changing the resolution and addition of noise. In real world observations, normalization would be the final step in data processing as in (a, b, c) and these fluctuation functions are in excellent agreement with those in Fig. \ref{fig:CO_Noise}.}
    \label{fig:Norm}
\end{figure*}
%}
%\bibliography{ExoRefs}

\section{Continuum Normalization}\label{Sec:normalized}
Due to change in observational conditions such as variation in cloud cover, the observed flux across wavelengths in spectral data is normalized to reduce such observation effects. This normalization acts as an additional filter on the magnitude of the observed flux. But it should also be noted that this normalization may interact with the stellar feature under study, since the timescale over which normalization is performed may be larger than the timescale of the stellar feature itself and thus mask any such features, e.g., stellar spots and faculae. Thus, while for a planet orbiting a star the normalization would mask out only the observational noise, for stellar spots and faculae, depending on their dynamic timescales, normalization can also act to mask the intrinsic dynamics. Fig. \ref{fig:Norm} shows the fluctuation functions for a spot (a), facula (b) and a planetary RV signal (c). Here, the normalization was performed as the final step in data processing, as would be done in actual observations. We also see that these fluctuation functions in Fig. \ref{fig:Norm} are in excellent agreement with those in Fig. \ref{fig:CO_Noise}.

\end{document}